\documentclass{tlp}
\usepackage{epsf}
\usepackage{multicol}
\usepackage{amssymb}
\usepackage{graphicx}
\usepackage{enumerate}
\usepackage{url}
\usepackage{aopmath}

\title[A Design and Implementation of the Extended Andorra Model]
      {A Design and Implementation of the \\ Extended Andorra Model\footnote{In memory of Ricardo Lopes, the main author of the YAP BEAM implementation.}}

\author[Ricardo Lopes, V\'{\i}tor Santos Costa and Fernando Silva]
       {RICARDO LOPES \hspace{0.3cm} V\'{I}TOR SANTOS COSTA  \hspace{0.3cm} FERNANDO SILVA\\
         CRACS-INESC Porto LA \& Faculdade de Ciencias, Universidade
         do Porto\\
         Rua do Campo Alegre 1021, 4169-007 Porto, Portugal\\
         \email{\{vsc,fds\}@dcc.fc.up.pt}
       }

\pagerange{\pageref{firstpage}--\pageref{lastpage}}

\submitted{22 March 2006}
\revised{3 July 2009}
\accepted{22 October 2010}

\begin{document}

\label{firstpage}

\maketitle

\begin{abstract}
  Logic programming provides a high-level view of programming, giving
  implementers a vast latitude into what techniques to explore to
  achieve the best performance for logic programs. Towards obtaining
  maximum performance, one of the holy grails of logic programming has
  been to design computational models that could be executed
  efficiently and that would allow both for a reduction of the search
  space and for exploiting all the available parallelism in the
  application. These goals have motivated the design of the Extended
  Andorra Model, a model where goals that do not constrain
  non-deterministic goals can execute first.

  In this work we present and evaluate the Basic design for Extended
  Andorra Model (BEAM), a system that builds upon David H. D. Warren's
  original EAM with Implicit Control. We provide a complete
  description and implementation of the BEAM System as a set of
  rewrite and control rules. We present the major data structures and
  execution algorithms that are required for efficient execution, and
  evaluate system performance.

A detailed performance study of our system is included.  Our results
show that the system achieves acceptable base performance, and that a
number of applications benefit from the advanced search inherent to
the EAM.

\end{abstract}

\begin{keywords}
Logic Programming, Implementation, Extended Andorra Model.
\end{keywords}

\section{Introduction}

Logic programming~\cite{Lloyd87} (LP) relies on the idea that
computation is controlled inference. LP provides a high-level view of
programming where programs are fundamentally seen as a collection of
statements that define a model of the intended problem. Queries may be
asked against this model, and answers will be given through a proof
procedure, such as refutation.  Prolog~\cite{Col93} is the most
popular logic programming language.  Prolog relies on SLD
resolution~\cite{Hill74}, and uses a straightforward left-to-right
selection function and depth-first search rule. This computation rule
is simple to understand and efficient to implement but, unfortunately,
it is not the ideal rule for every logic program. It is well known
that for many programs the left-to-right selection function is not
effective at constraining the search space and in the worst cases can
lead to looping. Often, these limitations lead Prolog programmers to
convoluted and non-declarative programs.


Ideally, one would like novel computational models for LP to achieve
the following two goals, in order of priority~\cite{War90}:
\begin{itemize}
\item \emph{Minimum number of inferences}: by trying never to repeat
  the same execution step (inference) in different locations of the
  execution tree.
\item \emph{Maximum parallelism}: by allowing goals to execute as
  independently as possible, and combining all solutions as late as
  feasible.
\end{itemize}

Both goals can be achieved through a variety of
techniques. \emph{Coroutining} and \emph{tabling} are nowadays widely
used to reduce the number of inferences performed by logic
programs~\cite{sicstus,yap,Eclipse,xsb-manual}. Coroutining allows goal
execution when all required arguments are bound. Tabling avoids
repeated computations of the same goal, and can be used to prevent infinite
loops. Several forms of parallelism, such as \emph{and-parallelism}
between goals, and \emph{or-parallelism} between alternatives, have
been exploited in LP, with excellent results~\cite{par-survey}.

One should observe that the two goals of minimal inferences and of
maximal parallelism are not independent. Indeed, work on concurrent LP
languages showed the strong interplay between coroutining and
and-parallelism~\cite{UedaMor,ghc02}. In the Basic Andorra Model
(BAM)~\cite{War88} coroutining between determinate goals (goals with
at most one valid alternative) constrains the search space and
generates and-parallelism, whereas the alternatives of
non-deterministic goals generate \emph{or-parallelism}. The Andorra-I
prototype~\cite{ppopp} demonstrated the approach to be practical and
effective. On the other hand, Andorra-I does depend on finding
determinacy. If determinacy cannot be found efficiently, there is no
benefit in using this model.

The Extended Andorra Model~\cite{War89} (EAM) lifts the main
restrictions in the BAM. The key ideas for
this model can be described as:
\begin{itemize}
\item Goals can execute immediately (in parallel) as long as they are
  deterministic or \emph{they do not need to bind external variables};
\item If a goal must bind external variables non-deterministically,
  the computation of this goal will \emph{split}.
\end{itemize}


The EAM provides a generic model for the exploitation of coroutining
and parallelism in LP, and motivated two main lines of research.

One approach was followed by Haridi, Janson and researchers at SICS
who concentrated on the AKL~\cite{JansonHSLP91}, the Agents Kernel
Language, based on the principle that the advantages of the EAM
justified a new programming paradigm that could subsume both
traditional Prolog and the concurrent logic languages.  AKL programs
are formed of guarded clauses, where the guard is separated from the
body through the sequential conjunction operator, cut, or commit.
AKL systems
obtained acceptable performance, both in sequential and parallel
implementations, such as Penny~\cite{Penny,PennySimICS}, but the
language was not actively supported.  Instead, the AKL researchers shifted
their interest to Oz~\cite{Oz}. This language provides some of the
advantages of LP such as the logical variables, and of AKL such as
encapsulation, but makes thread programming and search control fully
explicit.

In contrast, David H. D. Warren and researchers at Bristol
concentrated on the \emph{Extended Andorra Model with Implicit
Control}~\cite{War90}, where the goal was to apply the EAM as a
technique to achieve efficient execution of logic programs, with
minimal programmer effort. Gupta's proof-of-concept
interpreter~\cite{gupta-warren91} showed the need for further research
on the EAM, and presented new concepts such as \emph{lazy copying} and
\emph{eager producers} that give finer control over search and improve
parallelism. Gupta and Pontelli later experimented with an extension
of dependent and-parallelism that provide some of the functionality of
the EAM through parallelism, EDDAP~\cite{Gupta:1997:EDD}. EDDAP shows how
the EAM ideas are important in parallel logic programming systems.

In this work we present the BEAM, an implementation of the Extended
Andorra Model for Logic Programs. Our research was motivated by the
original question of whether the EAM can be an effective mechanism for
the execution of logic programs, and this work extends the original EAM work by:
\begin{itemize}
\item Providing a complete description of an EAM kernel as a set of
  rewrite and control rules, and evaluating these rules through a
  prototype implementation~\cite{rslvscfds_clps03}. We call this
  kernel design the BEAM, Basic design for Extended Andorra
  Model~\cite{rslvscfds_padl01}. Sections~\ref{sec:concepts} and~\ref{sec:simplification_rules} present this contribution.
\item Studying how to take the best advantage of the EAM with the
  least programmer intervention~\cite{rslvscfds_padl04}.  In the
  spirit of Kowalski's original definition, and building upon Warren
  and Gupta's original work~\cite{gupta-warren91}, we experimented
  with different approaches to exploiting control and contrast them to
  the guard-style approach used in AKL. Section~\ref{control_for_beam} presents this contribution.
\item Exploring novel implementation techniques for the EAM,
  including efficient support for deterministic
  computations~\cite{rslvscfds_iclp03} and efficient memory
  management~\cite{rslvsc_padl05}. Sections \ref{sec:architecture} and \ref{sec:memory} presents this contribution.
\end{itemize}

Our results show that the system achieves acceptable base performance,
and that a number of applications benefit from the advanced search
inherent to the EAM. Moreover, we show that implicit control can be in
fact quite effective for a sizeable number of applications, and that
simple annotations can contribute to further improvements with little
programmer effort.


This paper is organised as follows. Section~\ref{sec:concepts} presents
the main BEAM concepts, that fully specify an Extended Andorra Model
with implicit control. Next, in Section~\ref{sec:simplification_rules} we
propose a number of rules that simplify and optimize the BEAM
computational state. Section \ref{sec:architecture} shows the BEAM
implementation. Section~\ref{sec:memory} discusses memory management issues,
and Section~\ref{sec:emulator} focuses on emulator design. We evaluate
the performance of our system in Section~\ref{sec:performance}, and
finish with Conclusions and a Discussion of Related
Work. The reader is expected to have understanding of the key issues
in Logic Programming implementation, and in particular of the design of the
Warren Abstract Machine (WAM)~\cite{Warren83}.


\section{BEAM Concepts}
\label{sec:concepts}
A BEAM \emph{computation} is a sequence of rewriting operations
performed on And-Or Trees. And-Or Trees are trees of and-boxes and
or-boxes:

\begin{itemize}
\item An \emph{and-box} $\Delta$ represents a conjunction:

 \[ 
  [ \exists X_{1}, \ldots , X_{m}: \sigma~\&~ A_{1} ~\&~ \ldots \&~A_{n}~ ]~~(n\geq 0) 
\] 

  Each $A_i$ in the conjunction may be a literal $G_i$ or an or-box. Initially, an and-box 
  represents the body of a Horn clause and all $A_i$ are
  literals of the form $G_i$.

A variable is said to be \emph{local} to an and-box $\Delta$ when it
is scoped at $\Delta$.  The variables $X_1$ to $X_m$ represent the set of 
  variables local to the box $\Delta$. Initially, these variables are
  the variables occurring in and only in the body of the clause 
 $G_{1} ~ \&~ \ldots ~\&~G_{n}$.  

  The set $\sigma$ is a set of constraints. In this work, we shall
  focus only on Herbrand constraints, and we may refer to them as
  bindings.

  The \emph{environment} of an and-box consists of all variables local
  to the and-box and to every ancestor and-box. Variables local to an
  ancestor box are called \emph{external} to the current and-box. 

\item An \emph{or-box} $\Omega$ represents the matching clauses for a goal; each
  or-box contains a sequence of and-boxes $\Delta_{1}$ to $\Delta_{n}$.

  \[ \{\Delta_{1} \vee \ldots \vee \Delta_{n}\}~~(n\geq 0)\] 

  Each child and-box $\Delta_i$
  initially represents an alternative clause for a goal. 

\end{itemize}

A \emph{configuration} is an And-Or Tree describing a state of the
computation.  Given a literal $Q$, the query, a \emph{computation} is a sequence of
configurations starting at the initial configuration, and obtained by
successive applications of the rewrite rules defined next. The
\emph{initial configuration} is a single and-box such that:
\begin{itemize}
\item $X_1,\ldots,X_n = vars(Q)$
\item $\sigma = \emptyset$
\item the literal $Q$.
\end{itemize}

The constraints over the uppermost and-box(es) on the final
configuration are called the \emph{answer(s)}.

A goal, or literal, is said to be \emph{deterministic} when the
corresponding or-box has at most one and-box. Otherwise it is said to
be \emph{non-deterministic}.

An and-box $\Delta$ is said to be
\emph{suspended} if only the splitting rule, defined in
section~\ref{sec:beam_rules}, applies to $\Delta$ and if there is at
least one variable $X$ such that binding $X$ will allow applying a
different rule to $\Delta$. 

\subsection{Rewrite Rules}
\label{sec:beam_rules}

Execution in the EAM proceeds as a sequence of rewrite operations on
configurations. The BEAM's rewrite rules are based on David H.\ D.\
Warren's proposal~\cite{War90}. They have been designed to
allow for efficient implementation. In the following we use square
brackets to represent an and-box $\Delta$, curly brackets to represent an
or-box $\Omega$, and the symbol $G$ to
refer an unfolded literal (sub-goal), and the symbols $A$ and $B$ to
refer a sequence with literals and or-boxes. We present the rewrite rules
both graphically and textually. Graphically, the leftmost box is the
original configuration and the rightmost box the transformed
configuration. Textually, the configuration above the line is
transformed into the configuration below the line.

The BEAM rewrite rules are as follows:

\paragraph{Reduction}

This rule resolves a goal $G$ in an and-box against the heads of all
  clauses defining the procedure for $G$. The rule always creates a new
  or-box and one and-box for each clause that unifies with the goal.
  Each and-box $i$ is initialised with the most general unifier
  between the clause's head and the goal, $\sigma_i$, and with the set
  of existential variables in the clause, ${\cal Y}_i$. $A$ and $B$
  denote conjunctions of goals.

\[ \begin{array}{lc}
 & [ \exists {\cal X}: \sigma \& A \& G \& B ] \\
(Reduction) & \longrightarrow \\
& { [ }
\exists {\cal X}: \sigma \& A \& \left\{
\begin{array}{c}
{[} \exists {\cal Y}_{1}: \sigma_{1}~ \&~
 G_{11} \& \ldots \&~ G_{1k} ] \\
\lor 
  \ldots \lor \\
{[} \exists {\cal Y}_{n}: \sigma_{n}~ \&~  G_{n1} \& \ldots \&~
  G_{nk} ]
\end{array}
 \right\}
 \& B ]
\end{array}
\]

 \begin{figure}[htbp]
    \centerline{
      \includegraphics{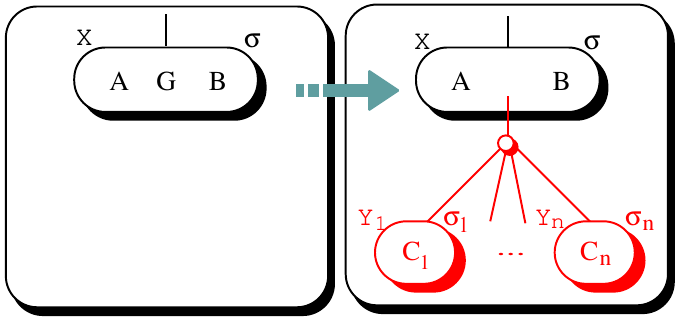}
    }
    \caption{BEAM reduction rule. Each $C_i$ represents the body of a clause $G_{i1}~\& \ldots \&~ G_{ik}$}
    \label{fig:beam-reduction}
  \end{figure}
  \noindent
  
  Figure \ref{fig:beam-reduction} shows how resolution expands the
  tree. Notice that the new variables created by the rewrite rule are
  guaranteed to be standardised apart. Also, the reduction rule just
  unfolds goal $G$, no constraint propagation is performed from below
  to above, even if the or-box has a single child.

\paragraph{Promotion} This rule promotes the variables and constraints from an
  and-box $\Delta$ to the closest ancestor and-box $\Delta'$:

\[ \begin{array}{lc}
& [\exists {\cal X}: \sigma~\&~A~\&~ \{[\exists {\cal Y}: \theta~\&~W]\}~\&~B] \}\\
(Promotion) & \longrightarrow \\
& { [}\exists {\cal X},{\cal Y}:
    \sigma\theta~\&~A~\&~\{[W]\}~\&~B]
\end{array}
\]

\begin{figure}[htbp]
    \centerline{ 
    \includegraphics{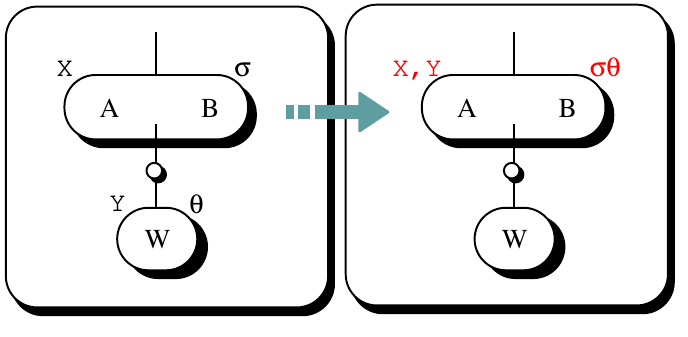} }
    \caption{BEAM promotion rule.}
    \label{fig:beam-promotion}
  \end{figure}

  The BEAM allows promotion only if $\Delta$ is the single alternative
  to the parent or-box, as illustrated in
  Figure~\ref{fig:beam-promotion}. The box $\Delta$ is represented by
      the round box that contains goal W and $\Delta'$ is represented
      by the round box that contains A and B. $\sigma\theta$ is the
      composition of constraints $\sigma$ and $\theta$.

  Promotion follows Warren's EAM rule in that it propagates results
  from a local computation to the level above.  However, promotion in
  the BEAM does not simplify the structure of the tree, in contrast
  with the original EAM and with AGENTS~\cite{JaMon92}.


\paragraph{Propagation} This rule allows us to propagate a constraint $\sigma_i$
  from an and-box to another and-box in the subtree below. This rule is thus symmetrical
  to the promotion rule.
   \[ 
 \begin{array}{lc}
& {[}\exists {\cal X},{\cal Z}: \sigma\& A ~\& ~
     \{ \ldots \vee [\exists {\cal Y}: \theta \& W]
   \vee \ldots \} ~\& B ]~~ \land  ~~\sigma_i \in \sigma\\
(Propagation) & \longrightarrow \\
&    {[}\exists {\cal X},{\cal Z}: \sigma \& A ~ \& ~\{
    \ldots \vee [\exists {\cal Y}:
  \theta\sigma_i \& W] \vee \ldots \}~ \& B] 
\end{array}
\]

   \begin{figure}[htbp]
     \centerline{
       \includegraphics{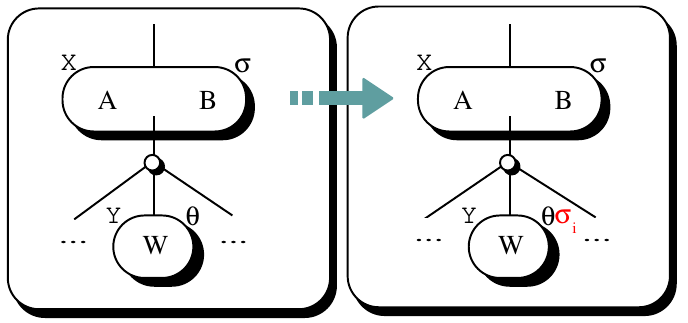}
     }
     \caption{BEAM propagation rule.}
     \label{fig:beam-propagation}
   \end{figure}

   Figure~\ref{fig:beam-propagation} shows how the
   propagation rule makes the
   constraint $\sigma$ available to the underlying
   and-boxes. Together, the promotion and propagation rules allow us
   to propagate bindings through the and-or tree. 
   
   \paragraph {Splitting} This rule is also known as
   \emph{non-determinate promotion}. The rule distributes a
   \emph{cut-free conjunction} across a disjunction, in a way similar to the original
   EAM's forking
   rule~\cite{War89}.

\[ \begin{array}{lc}
& [ \exists {\cal X}: \sigma~ \&~ A~\& \{ \Delta_1 \lor \ldots \lor
\Delta_i  \lor \ldots \lor \Delta_n\}~ \&~ B] \\
(Splitting) & \longrightarrow \\
& \{ [\exists {\cal X}: \sigma~ \&~ A~ \&~  \{ \Delta_i \}~ \&~ B]~\lor\\ 
& { [}
 \exists {\cal X}: \sigma~ \&~ A~ \&~ \{  \Delta_1 \lor \ldots \lor \Delta_{i-1} \lor  \Delta_{i+1}
   \lor \ldots \lor \Delta_n \} ~ \& ~ B] \} 
\end{array}
 \]

\begin{figure}[htbp]
\centerline{ 
\includegraphics{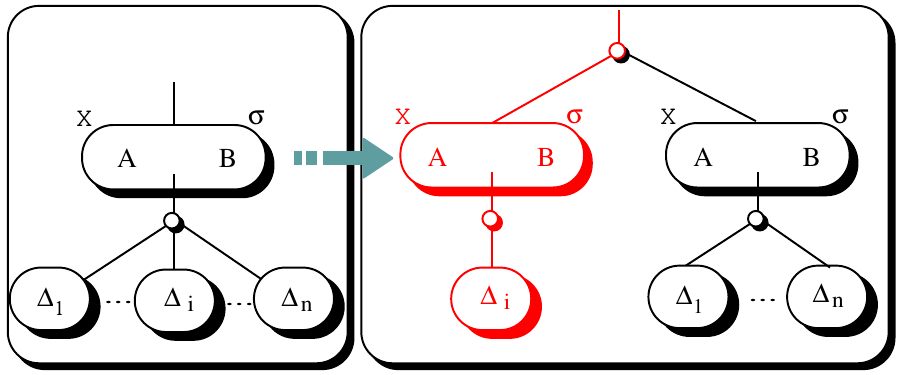} }
\caption{BEAM splitting rule.}
\label{fig:beam-splitting}
\end{figure}
\noindent As stated above, the parent and-box may not include a
pruning operator as its direct element.

In contrast to the previous rules, splitting duplicates goals in the
tree. It is therefore the most expensive rule in the BEAM, and as
discussed next, the control strategy in the BEAM tries to delay
application of splitting as much as possible.


\section{Simplification Rules }
\label{sec:simplification_rules}

We have presented the main rules that allow us to create and expand
the tree, and to propagate bindings. An actual implementation must be
able to simplify the And-Or Tree in order to propagate success and
failure and in order to recover space by discarding boxes. The BEAM
therefore includes additional simplification rules that generate
compact configurations and allow us to optimize the computation
process. These are:

\paragraph{Success-Propagation:}
  The original EAM does not explicitly provide a notion of successful
  computation, meaning  a computation that has completed execution and that may now be
  discarded. The following rules identify success situations in the
  BEAM and allow the propagation of success towards the upper boxes.
  To implement these rules we use the notion of a \emph{true-box}: an
  and-box is called a \emph{true-box} when all the local computations have been
  completed and the and-box does not impose constraints on external
  variables: 
  
\[ [\exists {\cal X}: \emptyset ] \equiv \mathtt{true} \]
  
  Note that the and-box might initially have had constraints on external
  variables, but those constraints have left the and-box after
  application of the promotion rule.  We say that a success
  occurs when we find a true-box. 
    
   \[  \begin{array}{lc}
(Success-Propagation) &[ \exists {\cal X}: \sigma ~\& ~A~\& ~ \{ \mathtt{true}\} ~\& ~ B ]
\longrightarrow 
 [ \exists {\cal X}: \sigma ~\& ~A~\& ~B ]
\end{array}
\]

\begin{figure}[htbp]
\centerline{
\epsfxsize=11.5cm
\includegraphics{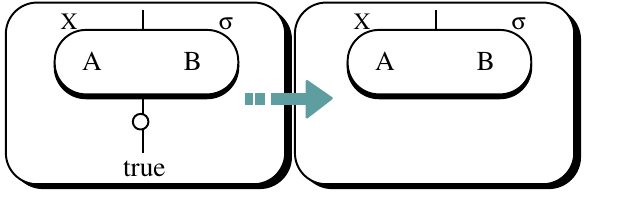}
}
\caption{BEAM Success-Propagation rule.}
\label{fig:beam-or-simp}
\end{figure}

If an or-box contains a unique alternative which succeeds, the or-box
has succeeded and can be discarded (see
Figure~\ref{fig:beam-or-simp}).  True-boxes may also be used to
achieve implicit pruning, as discussed below in
section~\ref{sec:imp_prun}.

  



  
  The ultimate goal of a BEAM computation is to reduce all and-boxes
  to true-boxes so that the initial and-box can itself be simplified.

\paragraph{Leftmost-Failure Propagation} The operation symmetric to the propagation of
  success is the propagation of failure. It is quite important to
  identify failed computations in order to allow the propagation of
  failure towards the upper boxes and in order to recover space. Again, the
  basic EAM design does not contain explicit rules for
  failure propagation. First, we define a \emph{fail-box} as an empty or-box:

  \[ \{\} \equiv  \mathtt{false} \]

  Failure can
  then be propagated by discarding the parent and-box:
 
   \[   
\begin{array}{lc}
&\{ \ldots \lor  \Delta_{i-1} \lor [ \exists {\cal X}: \sigma \&
~\mathtt{false}~  \&~ B ] \lor \Delta_{i+1}\lor 
   \ldots \} \\
(Leftmost-Failure) &\longrightarrow \\
&  \{ \ldots \lor  \Delta_{i-1} \lor \Delta_{i+1} \lor
   \ldots \} 
\end{array}
    \]

\begin{figure}[htbp]
\centerline{
\epsfxsize=9cm
\includegraphics{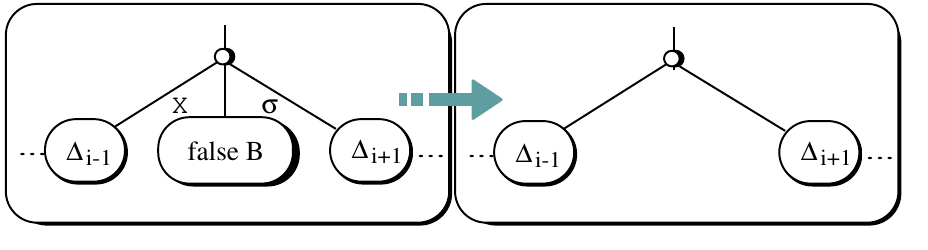}
}
\caption{BEAM Leftmost-Failure propagation rule.}
\label{fig:beam-false-and}
\end{figure}

Notice that the rule states that if the \emph{first} or leftmost goal of an
and-box fails, then the and-box has failed
    (see figure~\ref{fig:beam-false-and}).  A non-leftmost version of
    this rule for simplification of and-boxes is presented in the
    context of our discussion on pruning in section~\ref{sec:imp_prun}.

 Failure propagation rules such as this one have priority over all the other rules
  to allow propagation of failure as soon as possible.
  
\paragraph{And-Compression} 
The last rule addresses propagation of deterministic computations by
discarding or-boxes that have a single leaf:
  
  \[\begin{array}{lc}
& [\exists {\cal X}: \sigma~\&~ A~ \&~ \{ [ \exists {\cal Y}: \theta \& W ] \}~\&~ B]\\
(And-Compression) &\longrightarrow \\
& {[} \exists {\cal X},{\cal Y}: \sigma \& \theta~ \&~ A~ \&~ W~ \&~ B] 
\end{array}
\]

\begin{figure}[htbp]
\centerline{
\includegraphics{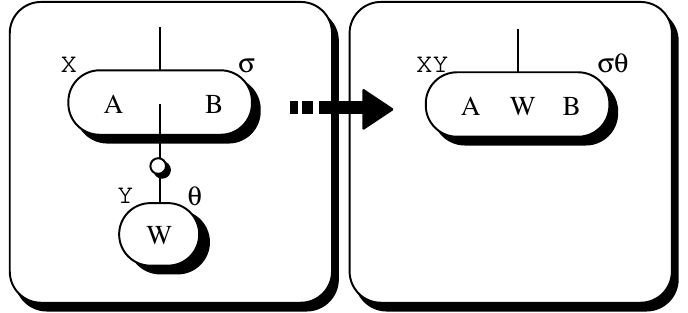}
}
\caption{BEAM And-Compression rule.}
\label{fig:beam-and-compression}
\end{figure}

This rule removes a nesting of two and-boxes by promoting the inner
  and-box to the outer and-box (see figure~\ref{fig:beam-and-compression}).  
  It thus complements the promotion rule by allowing the BEAM to
  discard structure. The BEAM does not apply this rule if there is a pruning operator such as
  \texttt{cut} in the inner box (see section \ref{sec:cut} for more
  details on pruning).

  Doing and-compression has several benefits. First, the BEAM can
  recover memory immediately. Second, the compressed tree becomes
  smaller and easier to manage. Last, there is less information
  to duplicate when performing splitting.

\subsection{Implicit Pruning}
\label{sec:imp_prun}

The BEAM implements two major simplifications that improve the
search space by pruning logically redundant branches, \emph{even when they
are not leftmost}. The two rules
are symmetrical: one is concerned with failed boxes, the
other is concerned with successful boxes.

\paragraph{False-in-And} this simplification rule removes an and-box
that is parent to a false box.

   \[
\begin{array}{lc}
&\{ \ldots \lor  \Delta_{i-1} \lor [ \exists {\cal X}: \sigma ~\&~ A~~
      \&~ \mathtt{false}~ \&~ B ] \lor \Delta_{i+1}\lor 
   \ldots\} \\
(False-in-And) & \longrightarrow \\
&  \{ \ldots \lor  \Delta_{i-1} \lor \Delta_{i+1} \lor
   \ldots \} 
\end{array}
    \]

\begin{figure}[htbp]
\centerline{
\epsfxsize=9cm
\includegraphics{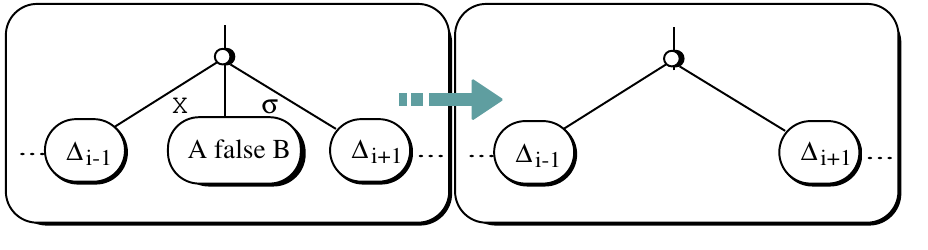}
}
\caption{BEAM False-in-And simplification rule.}
\label{fig:beam-false-inand}
\end{figure}

  If a failure occurs at any point of the and-box, the and-box is
  removed and failure is propagated to the upper boxes
  (see figure~\ref{fig:beam-false-inand}). This rule can
  be considered a generalisation of the failure propagation strategies used
  in Independent And-Parallel systems~\cite{HerNasr86}.
  
\paragraph{True-in-Or} This simplification rule removes an or-box that is
  parent to a true-box.

\[ 
\begin{array}{lc}
&[ \exists {\cal X}: \sigma ~\&~
     A~ \& \{ \ldots \lor~\mathtt{true}~ \lor \ldots \} \&~B ] \\
(True-in-Or) & \longrightarrow \\
& {[} \exists {\cal X}: \sigma ~\&~A~\&~B ] 
\end{array}
\]

\begin{figure}[htbp]
\centerline{
\includegraphics{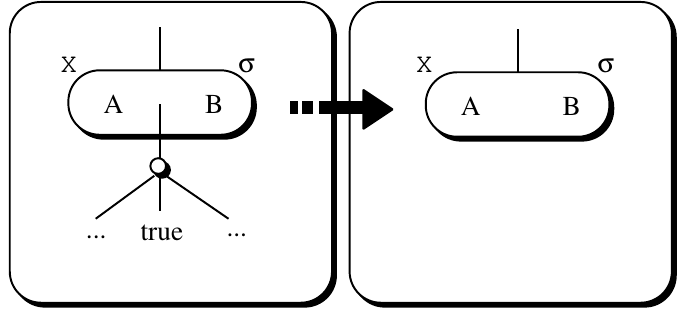}
}
\caption{BEAM True-in-Or simplification rule.}
\label{fig:beam-true-inor}
\end{figure}

This form provides implicit pruning of redundant branches in the
search tree (see figure~\ref{fig:beam-true-inor}), and it generalizes
XSB's work on early completion of tabled
computations~\cite{Sagonas-PhD,Sagonas-98}.
  
\paragraph{Implicit Pruning} These two rules can provide substantial
pruning. In the absence of side effects, the presence of a true-box in
a branch of an or-box should allow immediate pruning of all the other
branches in the or-box. In an environment with side-effects, the user
may still want the other branches to execute anyway. To guarantee
Prolog compatibility, the BEAM by default only allows the true-in-or
simplification and the false-in-and simplification for the leftmost
goal.  Alternatively, one could do compile-time analysis to disable
the true-in-or and false-in-and optimization for the boxes where some
branches include builtins calls with
side-effects\cite{andprolog,iclp91p}.

\subsection{Explicit Pruning}
\label{sec:cut}

The implicit pruning mechanisms we provide are not always sufficient
for controlling the search space. The BEAM therefore supports two
explicit pruning operators: cut (\textbf{!}) and commit (\textbf{$|$})
prune alternatives clauses for the current goal, plus alternatives for
all goals created for the current clause. Cut only prunes 
branches that appear to the right of the current branch, commit can
prune both to the left and to the right.  Both operators disallow goals to their right from
exporting constraints to the goals to the left, prior to execution.
After the execution of a cut or commit, all and-boxes to the right of the
and-box containing the cut operator are discarded and their constraints on external
variables are promoted to the current and-box.

\begin{figure}[htbp]
\centerline{ 
\includegraphics{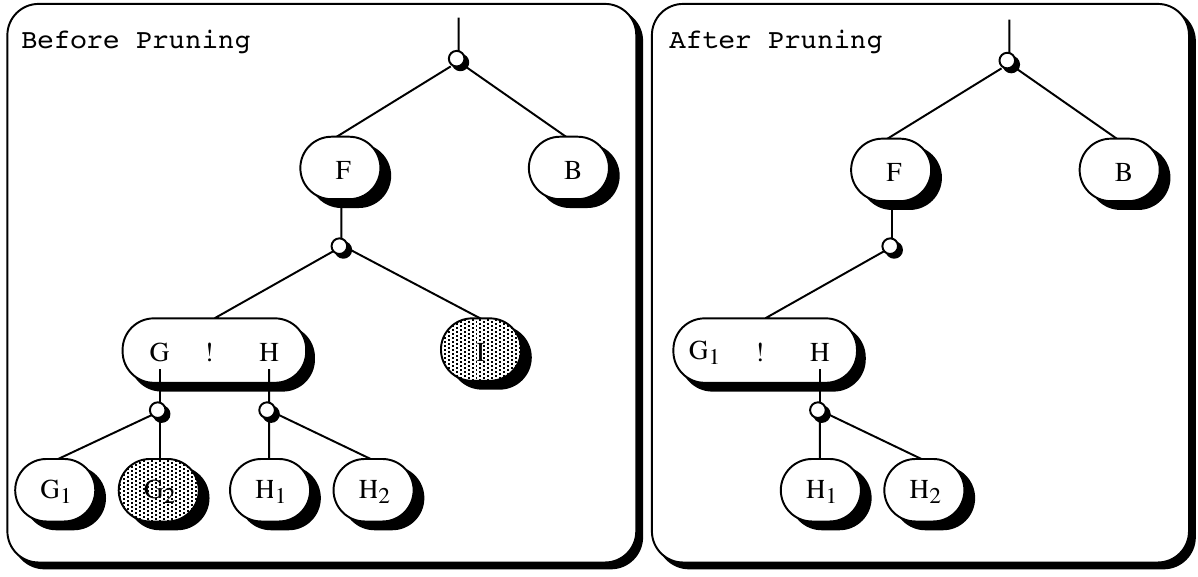} }
\caption{Cut scope.}
\label{fig:cut_example}
\end{figure}

Figure~\ref{fig:cut_example} gives an example of explicit pruning for
the program (we label the clauses of g/1 and h/1 to clarify the
figure):

\begin{verbatim}
a(X) :- f(X), b(X).    f(X) :- g(X), !, h(X).
a(X) :- b(X).          f(X) :- i(X).

G1: g(1).              H1: h(3).
G2: g(2).              H2: h(2).
\end{verbatim}

\noindent In this
example the and-boxes for the sibling clause \texttt{I} and for the
rightmost alternative \texttt{G2} will be discarded. Next, the constraints for
\texttt{G1} can be promoted to the and-box for \texttt{G, !, H}.

The \emph{cut} rule can be written as follows:
\[
\begin{array}{lc}
&\{[\exists {\cal X}: \theta_{1} \& \{[\exists {\cal Y}: \theta_{2} ] \vee \ldots \}
\&~\mathtt{!}~ \&~ A] \vee \ldots \} \\
(Cut)~~~~~~~~~~~~~~ &\longrightarrow \\

&  \{ [\exists {\cal X},{\cal Y}: \theta{_1}\theta{_2} \& A] \}
\end{array}
\]
\noindent
and the \emph{commit} rule as:
\[
\begin{array}{lc}
&\{\ldots \lor~[\exists {\cal X}: \theta_{1} \& \{\ldots \vee [\exists  Y: \theta_{2} ] \vee \ldots \} \&~\mathtt{|}~ \&~ A] \vee \ldots \} \\
(Commit) & \longrightarrow \\
&  \{ [\exists {\cal X},{\cal Y}: \theta{_1} \theta{_2} \& A] \}
\end{array}
\]

The conjunction of goals in the clause to the left of the cut or
commit is called its \emph{guard}. Our rule
states that cut only applies when the goals in the guard have been
unfolded into a configuration of the form:
\[ \{[\exists {\cal X}: \theta_{2} ] \vee \ldots \} \]

\noindent that is, when the leftmost branch of the guard has been completely
resolved. 

  


We next discuss in more detail the issues in the design of explicit
pruning for the BEAM.
In the following discussion we refer mainly to cut, but similar
arguments apply to the commit operator.

\section{Control for the BEAM}
\label{control_for_beam}

The previous rewrite and simplification rules provide the basic
engine for the correct execution of logic programs. To this
engine, we must add control strategies that decide which step to
take and when.  Describing when each rule is allowed to execute does
not completely define the BEAM control strategy. It is also necessary
to define the priority for each rule so that if a number of
rewrite-rules match, one can decide which one to choose. Arguably, one
could choose the same rule as Prolog, and clone Prolog execution.  The
real power of the EAM is that we have the flexibility to use different
strategies. In particular, we will try to find one that minimises the search space. The key ideas are:
\begin{enumerate}
\item Failure propagation rules have priority over all the other
  rules, so that failure propagates as fast as possible.
\item  Success propagation and and-compression rules should always be done
  next, because they simplify the tree.
\item Promotion and propagation rules should follow, because their
  combination may force some boxes to fail. 

  The BEAM favors two types of reductions. First, deterministic
  reductions, do not create or-boxes, and thus will never lead to
  splitting. Second, reductions that do not constrain external
  variables can go ahead early.
\item Last, the splitting rule is the most expensive operation, and should be
  deferred as much as possible.
\end{enumerate}


This default scheme of the execution control in BEAM thus favours
deterministic rules: the simplification rules, the promotion rule and
the propagation rule. Their implementation is therefore crucial to the
system performance as we expect that most execution time in the EAM
will be spent performing deterministic reductions, or reductions that
do not constrain the external environment.


\subsection{Improving Deterministic Work}
\label{sec:determinacy}

Warren's EAM proposal states that an and-box should immediately
suspend when trying to non-deterministically constrain an external
variable whose scope is above the closest or-box. Unfortunately, the
original EAM rule may lead to difficulties. We next discuss two
examples, a small data-base, \texttt{parent/2}, and a Prolog
procedure, \texttt{partition/4}, given in Figure~\ref{fig:partition}.

\begin{figure}[htp]
\begin{verbatim}
parent(john, richard).     partition([X|L],Y,[X|L1],L2) :- X =< Y,
parent(john, mary).                 partition(L,Y,L1,L2).
parent(patrick, paul).     partition([X|L],Y,L1,[X|L2]) :- X > Y,
parent(patrick, susan).             partition(L,Y,L1,L2).
                           partition([],_,[],[]).
\end{verbatim}
\caption{Prolog's \texttt{parent/2} and \texttt{partition/4} predicates.}
\label{fig:partition}
\end{figure}

Consider the query \texttt{?- parent(X,mary)}. The query is
deterministic, as it only matches the second clause. Unfortunately, a
naive implementation of Warren's rule would not recognize the goal as
deterministic. Instead, all four clauses would be tried, as
\texttt{parent/2} would try to bind a value to \texttt{X}, and all
and-boxes would suspend.  

The same problem may happen with the query: \texttt{?- partition([4,3,5],2,A,B)}.  
Although the calls are deterministic, the BEAM would suspend when unifying \texttt{[X|L1]} to
\texttt{A}.  The suspension would eventually lead to splitting and
to poor performance.

AGENTS~\cite{SverkerThesis} addresses this issue by relying on the
guard operators to explicitly control when goals can
execute. Arguably, this should allow for the best execution.  On the
other hand, AGENTS performance may be vulnerable to user errors, and
Prolog programs need to be pre-processed in order to perform
well~\cite{bueher92}. Andorra-I addresses a similar problem through
its compiler.  Unfortunately, coding all possible cases of determinacy
grows exponentially~\cite{PalNai91}.  In the end, Andorra-I manages
code size explosion by imposing a limit on the combinations of
arguments that it tries~\cite{iclp91p}.  This solution becomes a
source of inefficiency as Andorra-I often has to execute the same
unifications twice: initially, when checking for determinacy, and
later, after committing to a clause.

To address this problem, we propose a different control rule to define
when a reduction should suspend. Reduction of an and-box cannot
proceed and should therefore \emph{suspend} if and only if:

\begin{enumerate}[(i) ]


\item unification of the head arguments constrains external variables, and, 
\item at least two clauses \emph{unify} with the current goal.
\end{enumerate}

The BEAM performs full unification first and then checks whether these
conditions hold. These provides a more aggressive
determinacy scheme than the one in Warren's EAM which leads to
suspension immediately when binding an external variable.  One
advantage is that our scheme is simpler to implement, since the
suspension may only occur at a fixed point of the code thus reducing
the number of tests one needs to make in order to determine whether
the current and-box should or should not suspend.

Condition \textbf{(ii)} assumes that we are able to detect which
clauses may match. In practice this is the province of the
\emph{indexing} algorithm. It is known that detecting determinacy is
in general NP-complete~\cite{PalNai91}. Therefore, for efficiency
reasons, the indexing algorithm will be a conservative approximation.


\paragraph{Deterministic-reduce-and-promote} 

Performance of many logic programs heavily depends on optimisations such as Last Call
Optimisation. In the best case, such optimisations allow
tail-recursive programs to execute with the same costs that iterative
programs would.

Both EAM and AGENTS create an and-box when performing reduction on
deterministic predicates. The newly created and-box is promoted
immediately afterwards because it is deterministic. The creation of
boxes that are immediately promoted is expensive, both in memory usage
and in time.

\begin{figure}[htbp]
\centerline{
\includegraphics{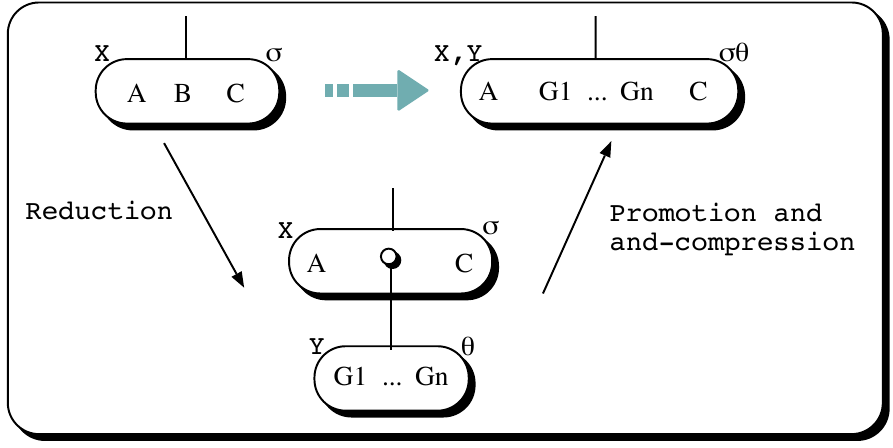}
}
\caption{BEAM Deterministic-reduce-and-promote rule.}
\label{fig:beam-det-red-promote}
\end{figure}

The BEAM addresses this problem through the
\emph{Deterministic-reduce-and-promote} rule. This rule allows for a
reduction to expand directly in the home and-box. More precisely,
whenever a deterministic goal $B$ is to be reduced to a single
alternative with goals $G_1,\ldots,G_n$, the reduction can take place in
the parent's and-box. Figure~\ref{fig:beam-det-red-promote} shows an
example of this rule.

    \[\begin{array}{lc}
(Deterministic-reduce & [ \exists {\cal X}: \sigma~ \&~ A~ \&~ B~ \&~ C] \\
~-and-promote) &\longrightarrow \\
&      {[} \exists {\cal X},{\cal Y}: \sigma\theta~ \&~ A~ \&~ G_1~ \& \ldots \&~ G_n~
\&~ C ] 
\end{array}
 \]

\noindent
As explained before, the BEAM cannot apply the
reduce-promote rule if there is a pruning operator, such as
\texttt{cut}, in the inner and-box.

\subsection{Control for Cut}

\begin{figure}[thbp]
\centerline{
\includegraphics{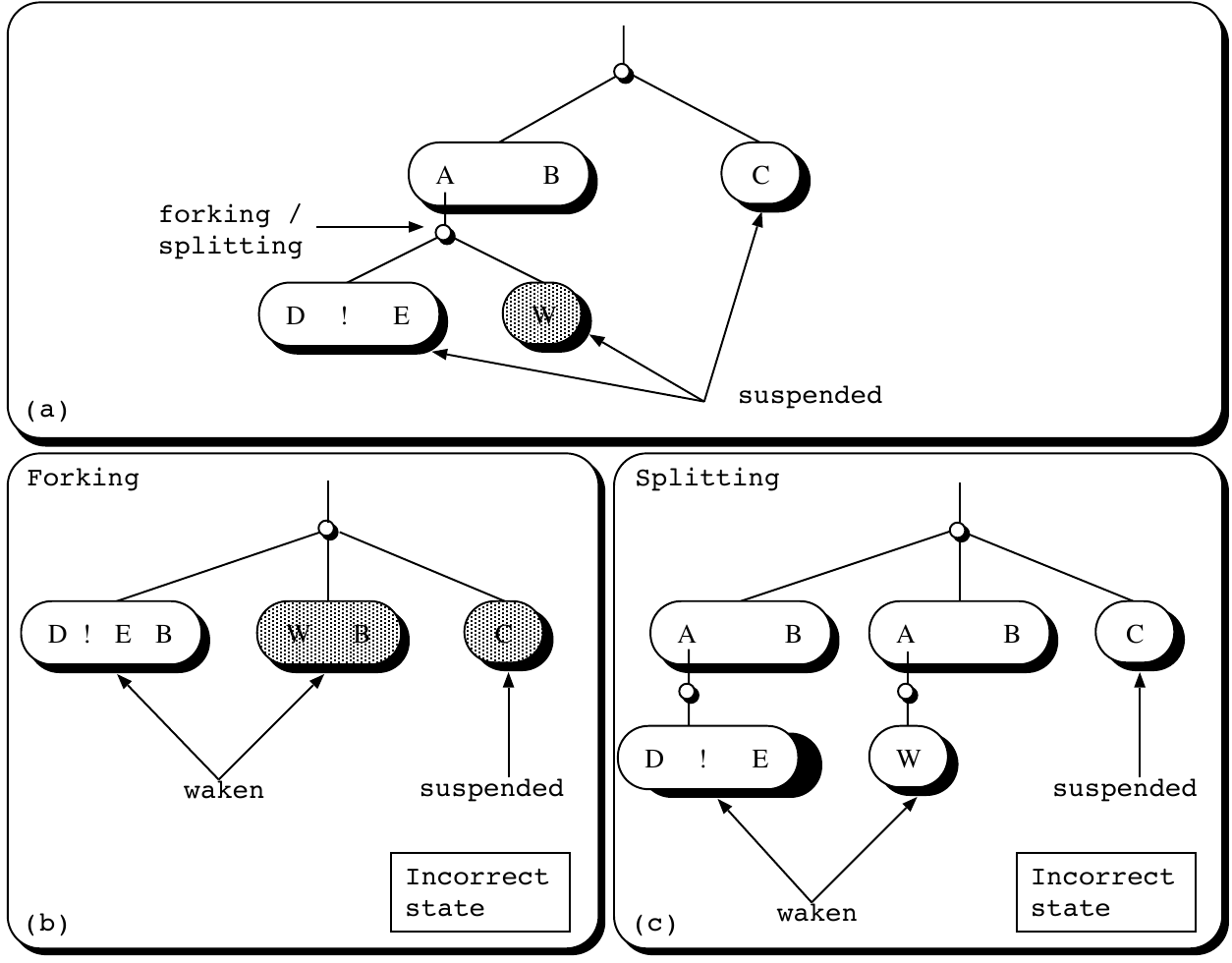}
}
\caption{Incorrect use of Fork/Splitting and Cut.}
\label{fig:cut_fork}
\end{figure}

Consider the example presented in Figure~\ref{fig:cut_fork}a, where
the lower-leftmost and-box contains a cut ({\tt D,!,E}). The and-box $
W$ is the only box within the cut scope.  Suppose that all boxes are
suspended and that the only available rule is
splitting. Unfortunately, splitting incorrectly allows the and-box $W$
to leave the scope of the cut (see figure~\ref{fig:cut_fork}c). An
alternative would be to resort to Warren's original forking rule, but
forking incorrectly allows the and-box $C$ to be deleted by the cut
(see figure~\ref{fig:cut_fork}b).

Please recall that the BEAM explicitly disallows the splitting of an
and-box containing a cut. A clause with a cut can continue execution
even if head unification constrains external variables (note that
these bindings may not be made visible to the parent boxes). In this
regard the BEAM is close to AGENTS~\cite{SverkerThesis}. The BEAM
differs from AGENTS in that goals to the right of the cut may also
execute before pruning: the cut does not provide sequencing.

\begin{figure}[htbp]
\centerline{
\includegraphics{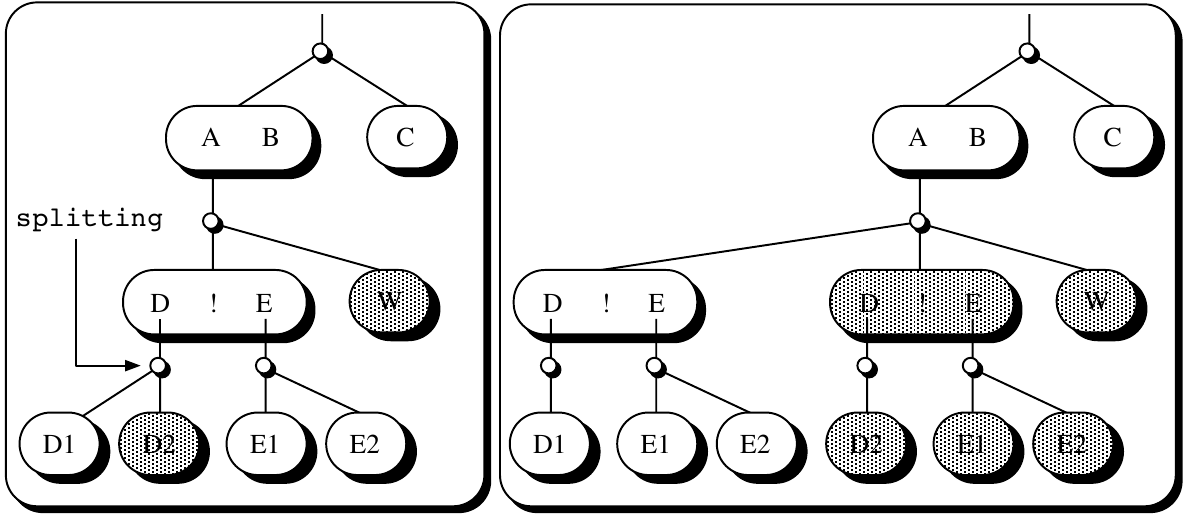}
}
\caption{Correct use of splitting and cut.}
\label{fig:cut_fork2}
\end{figure}

Splitting can be applied freely whenever the goals within the guard of
  the cut do not constrain external variables, but it 
  may not export constraints for variables external to the guard nor
  change the scope of the cut (see example in
  Figure~\ref{fig:cut_fork2}).

The following two rules are used to control cut execution:

\begin{itemize}
\item \emph{Early-pruning} - a cut can always execute immediately if it is the leftmost subgoal
  in the and-box and if the and-box does not have
  constraints.

\[ (Early-pruning)~~~~~\{ [\exists {\cal Y}: ~\mathtt{!}~
  \&~ \ldots] \vee W \} \longrightarrow
  \{ [\exists {\cal Y}: ~ \ldots] \} 
\]

Consider the example illustrated in Figure~\ref{fig:cut_leftmost}a.
Both alternatives to the goal {\tt a} suspended trying to bind the
external variable {\tt }. The first alternative to the goal {\tt b}
contains a \emph{quiet} cut~\footnote{a cut is quiet if its guard does
  not impose constraints on the caller's environment} that will be
allowed to execute since it respects the conditions described
previously: the alternative does not impose external constraints, and
the cut is the leftmost call in the and-box. Note that the resulting
execution here is close to the standard Prolog execution.

\begin{figure}[htbp]
\centerline{
\includegraphics{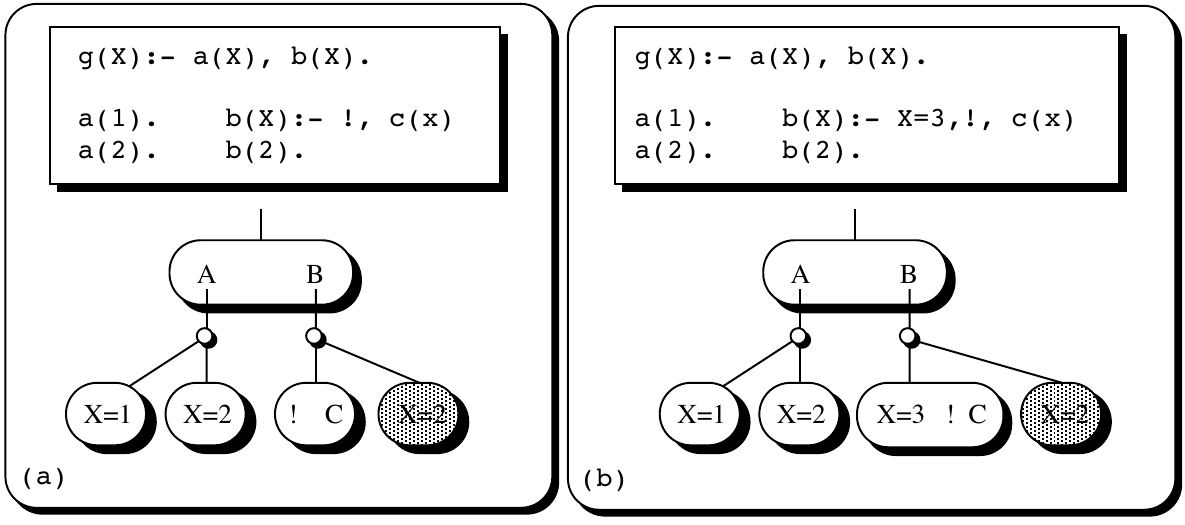}
}
\caption{Cut example.}
\label{fig:cut_leftmost}
\end{figure}

Figure~\ref{fig:cut_leftmost}b illustrates a different situation. In
this case, the cut would not be allowed to execute since the
alternative restricts the external variable {\tt X} to the value {\tt 3}.
Thus, the computation in this example would only be allowed to
continue after splitting on {\tt a}. After splitting, the values
{\tt 1} and {\tt 2} would be promoted to {\tt X} and thus make the first
alternative to {\tt b} fail.

\item \emph{Leftmost-pruning} - if an and-box containing a cut becomes leftmost in the tree, the
cut can execute immediately when all the calls before
it succeed (even if there are external constraints). 

\[ \begin{array}{lc} 
& [\exists {\cal X}_{1}: \theta_{1}\&\{ \ldots \lor \{ 
 [\exists {\cal X}_{n}: \theta_{n}\&\mathtt{!} \& D]
  \vee W \} \lor \ldots \} \&E ] \\
(Leftmost-pruning) &\longrightarrow \\
&  {[}\exists {\cal X}_{1}: \theta_{1}\& \{ \ldots \lor \{ 
  [\exists {\cal X}_{n}: \theta_{n}\&D] \} \lor \ldots \} \&E ] 
\end{array}
\]

For example, in Figure~\ref{fig:cut_leftmost_box} the cut is allowed to
execute immediately when \emph{X} succeeds even if there are external
constraints.

\begin{figure}[htbp]
\centerline{
\includegraphics{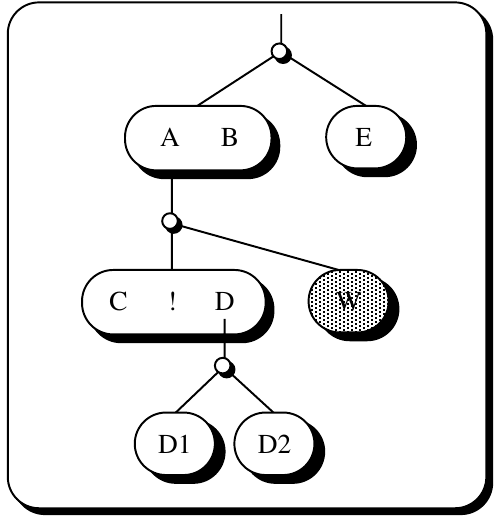}
}
\caption{Cut in the leftmost box in the tree.}
\label{fig:cut_leftmost_box}
\end{figure}

\end{itemize}

Allowing early execution of cuts will in most cases prune alternatives
early and thus reduce the search space.

\paragraph{Degenerate Parent Or-boxes}

\begin{figure}[htbp]
\centerline{
\includegraphics{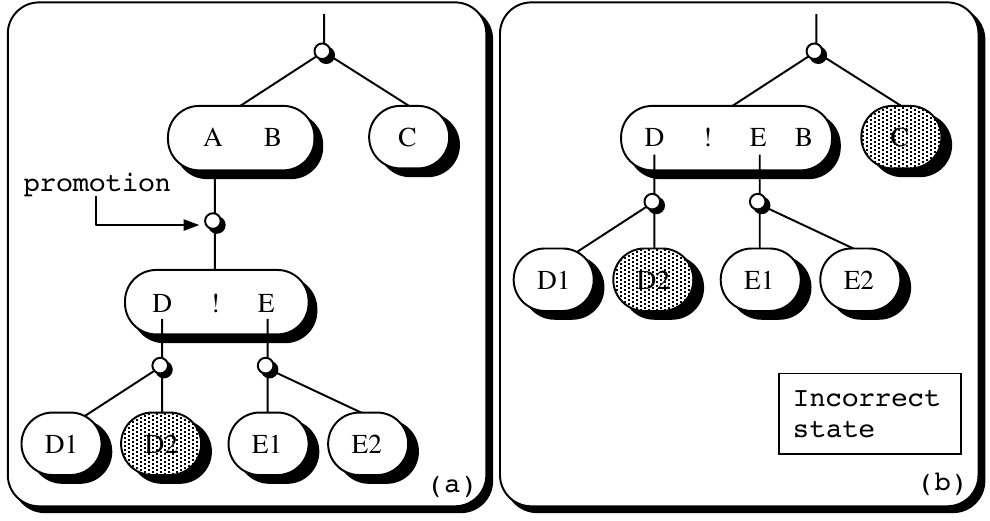}
}
\caption{Incorrect use of promotion and cut.}
\label{fig:cut_promotion}
\end{figure}

One interesting problem occurs when the parent or-box for the and-box
containing the cut degenerates to a single alternative. In this case,
promotion and and-compression would allow us to merge the two
resulting and-boxes. As a result, cut could prune goals in the
original parent and-box.  Figure~\ref{fig:cut_promotion} shows an
example where promotion of an and-box containing a cut leads to an
incorrect state as the and-box \emph{C} is in danger of being
removed by cut.

The BEAM disallows and-compression when the inner
and-box has a cut. Promotion of bindings is still allowed. Thus
deterministic constraints are still allowed to be exported.

\subsection{Non-Deterministic Work}
\label{sec:nondeterminacy}

Most programs have to perform splitting at some point. Deciding where
to apply the splitting rule is a fundamental issue for
EAM implementations.  We have considered three major extensions to the
default rule:
\begin{itemize}
\item Declare predicates as \emph{producers} and allow them to perform
  \emph{Eager-Forking}, that is, to do splitting as soon as they are
  called.  The idea was first proposed by Gupta and
  Warren~\cite{gupta-warren91}.  Intuitively, we declare goals to be
  producers if we expect them to produce, but not consume, bindings
  from other goals.  
  
  Eager-forking goes against the core idea of the EAM: doing
  determinate work first. On the other hand, it is simple to
  understand, it can
  increase parallelism, and in the cases where most alternatives in the
  producer fail quickly it can actually improve the search space. We
  have
  implemented
  eager-forking on the BEAM and we discuss some results in
  section~\ref{sec:performance}.

\item Allow the user to specify a
 boundary up to which we should check for splitting. The idea has
  appeared in many guises: mini-scopes in Gupta and
  Warren's simulator~\cite{gupta-warren91}, independent computations
  in Bueno and
  Hermenegildo~\cite{bueher92}.  Scoping is also quite important for a
  parallel implementation.

  AKL~\cite{SverkerThesis} introduces stability to allow early
  splitting. If only splitting applies to an and-box $\Delta$, the
  and-box $\Delta$ is
  said to be \emph{stable} if neither $\Delta$ nor any
  descendant and-box is suspended on variables external to $\Delta$. 
 All stable and-boxes can be split in parallel. Unfortunately,
 detecting stability is quite expensive.

  
\item The right-hand side of sequential conjunctions can 
  be evaluated only after the left-hand-side has succeeded. We do not allow
  sequential conjunction below the default (parallel) conjunction.
  This is sufficient to guarantee
  correct ordering for side-effects builtins such as \texttt{read/1}
  or \texttt{write/1}~\cite{iclp91p}, and allows a simpler
  implementation. \end{itemize}


In the next section we present the architecture
used to implement BEAM, namely the \emph{And-Or Tree Manager} and
the \emph{Abstract Machine}. 

\section{BEAM Implementation}
\label{sec:architecture}
\begin{figure}[htbp]
\centerline{
\includegraphics{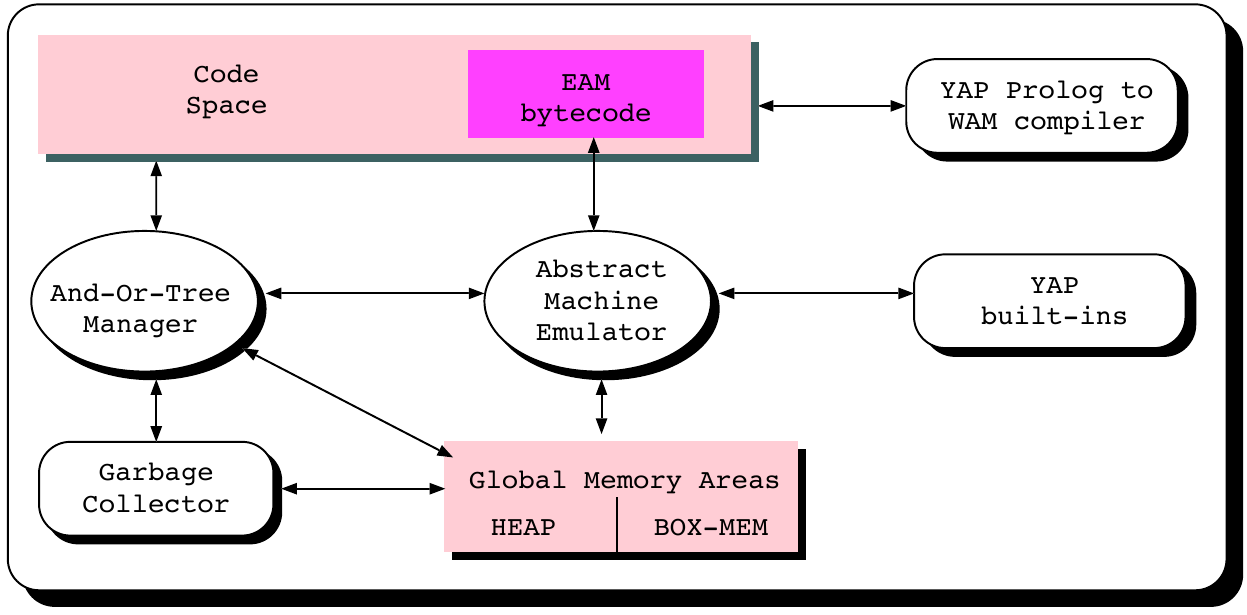}
}
\caption{Execution Model.}
\label{fig:execution_model}
\end{figure}

Figure~\ref{fig:execution_model} illustrates the architecture
organization for the BEAM execution model.  The BEAM was implemented
as an extension of the YAP Prolog system~\cite{yap-optim}. It reuses
most of the YAP \emph{compiler} and its \emph{builtin library}.  The
shadowed boxes show where the EAM stores data. The \emph{Code Space}
stores the data-base with the predicate/clause information, plus the
bytecode to be interpreted by BEAM's \emph{Abstract Machine Emulator}
(which we refer simply as \emph{Emulator}).  The \emph{Global Memory}
stores the And-Or Tree, and is further subdivided into the \emph{Heap}
and the \emph{Box Memory}.  The \emph{Box Memory} stores dynamic data
structures including boxes and variables. The \emph{Heap} holds Prolog
terms, such as lists and structures.  The \emph{Heap} uses term
copying to store compound terms and is thus very similar to the WAM's
\emph{Heap}.

There are a number of differences between the BEAM and the WAM. A
major difference is that the \emph{BEAM does not perform backtracking}. A \emph{Garbage 
  Collector} is thus necessary to recover space in the \emph{Heap}. We 
leave the details on memory management to section~\ref{sec:memory}.

The BEAM relies on two main components, shown as ovals in
Fig.~\ref{fig:execution_model}:
\begin{description}  
\item[Emulator:] runs WAM-like code to perform unification 
  and to setup boxes. Unification code is similar to the WAM. Control
  instructions follow a compilation scheme similar to the WAM but
  execute in a rather different way.
\item[And-Or Tree Manager:] applies the BEAM rewriting
  rules to the existing and-boxes and or-boxes until 
  WAM-like execution for the selected goal can start.
\end{description}

The \emph{And-Or Tree Manager} handles most of the complexity in the EAM. It
uses the \emph{Code Space} area to determine how many alternatives 
a goal has and how many goals a clause calls. With this information the
\emph{And-Or Tree Manager} constructs the tree and uses the EAM rewriting
rules to manipulate it. The Manager requests memory for the boxes from the
\emph{Global Memory Areas}. The \emph{Emulator} is called by the 
\emph{And-Or Tree Manager} in order to execute and unify the arguments of the 
goals and clauses.  As an example consider the clause:
\texttt{p(X,Y):- g(X), f(Y)}. When running this clause the
\emph{And-Or Tree Manager} transforms 
\texttt{p(X,Y)} into an and-box, and calls the \emph{Emulator} to create
the subgoals and or-boxes for
\texttt{g(X)} and \texttt{f(Y)}.
Control then moves to these subgoals,
and will return to the
\emph{And-Or Tree Manager} only if the and-boxes generated for these subgoals need to suspend.

The details on how BEAM stores the And-Or Tree, 
the design of the \emph{Emulator} and of the \emph{And-or Tree
manager} are described in more detail in the following sections.

\subsection{Or-Boxes}

\begin{figure}[htbp]
\centerline{
\includegraphics[width=12cm]{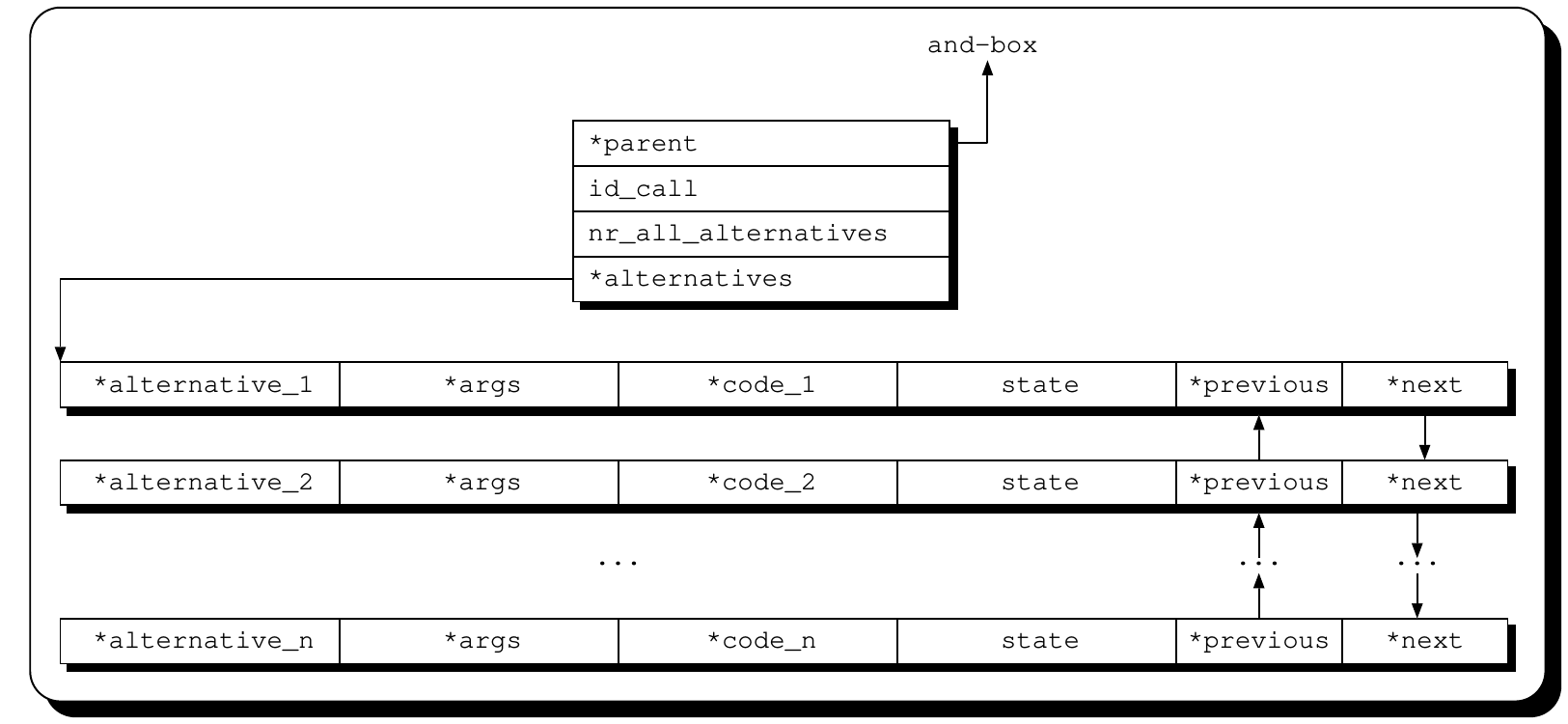}
}
\caption{Or-Box representation.}
\label{fig:orbox}
\end{figure}

Or-boxes represent open alternatives to a goal. Figure \ref{fig:orbox}
presents the structure of an or-box. Each or-box refers to its parent
and-box through the \texttt{parent} pointer, and to the sub-goal that
created the or-box through the \texttt{id\_call} field. The
field \texttt{nr\_all\_alternatives} counts the number of current
alternatives. Last, the box points to a list of \texttt{alternatives},
where each element in the list includes:
\begin{itemize}
\item a pointer to a corresponding and-box, \texttt{alternative},
  initially null; it is initialized only when the alternative is
  explored;
\item a pointer to the goal arguments, \texttt{args};
  The first alternative creates the arguments vector. The last
  alternative to execute, after performing head unification, recovers
  the {\tt args} vector as free memory. Each alternative needs a pointer
  to the {\tt args} vector because the \emph{and-compression}
  and the \emph{splitting} rules can join alternatives to different goals;

\item a pointer to the code for the alternative, \texttt{code}; and,
\item the \texttt{state} of the alternative. Initially, alternatives
  are in the \texttt{ready} state. They next move to the
  \texttt{running} state, from where they may reach the
  \texttt{success} or \texttt{fail} states, or they may enter the
  \texttt{suspend} state. Suspended alternatives will eventually move
  to the \texttt{wake} state. From \texttt{wake} state alternatives
  move to the \texttt{running} state again.  As an optimization, if no
  more sub-goals need to be executed, but the alternative is
  suspended, the alternative enters a special
  \texttt{suspended\_on\_end} state.
\end{itemize}

Note that if {\tt nr\_all\_alternatives} is one, then there is no need
to keep the or-box: the determinate promotion rule can be
applied to a single alternative. If {\tt nr\_all\_alternatives} equals
zero, the box has failed and we can perform \emph{fail propagation}.

\subsection{And-Boxes}
\begin{figure}[htbp]
\centerline{
\includegraphics[width=12cm]{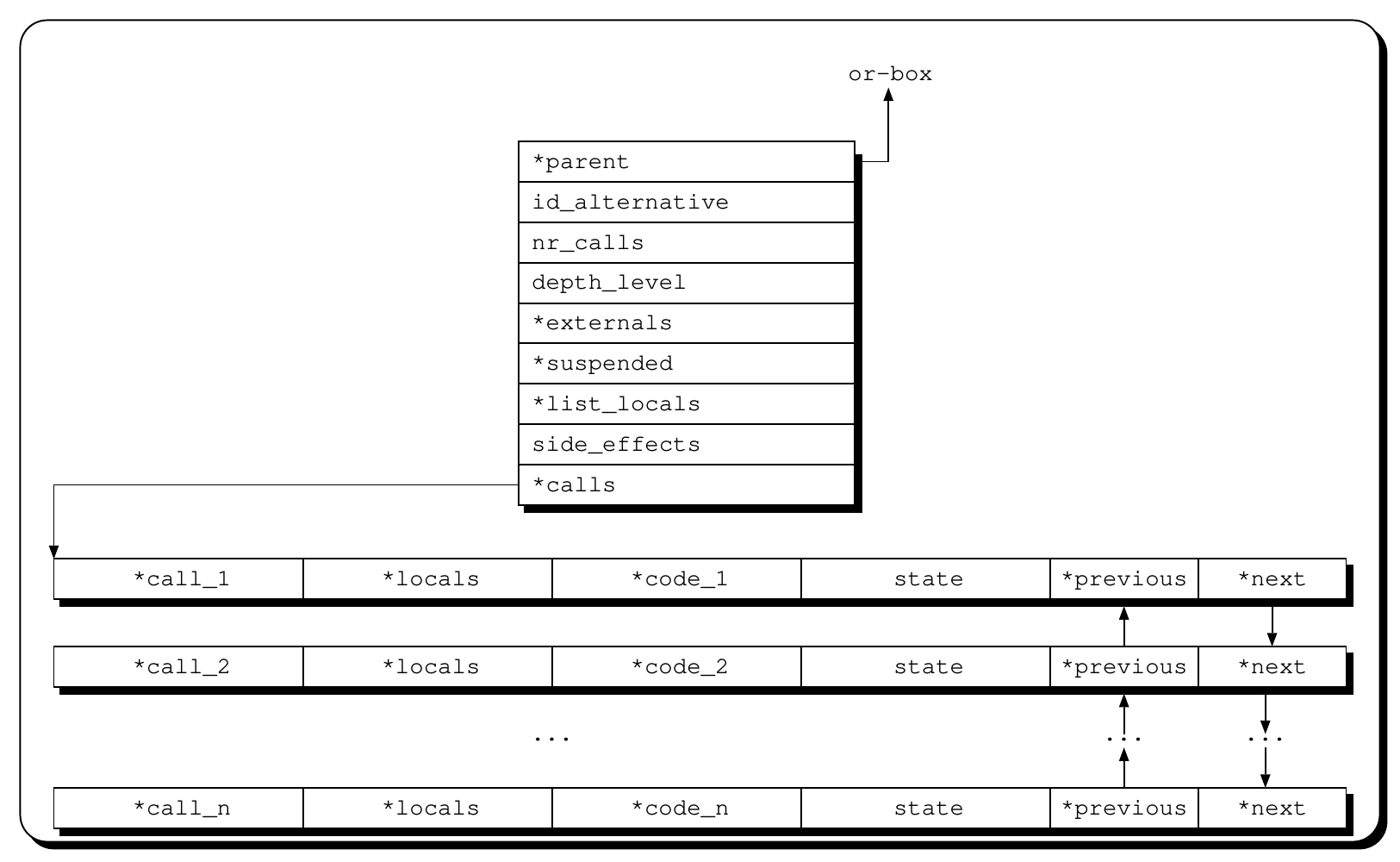}
}
\caption{And-Box representation.}
\label{fig:andbox}
\end{figure}

And-boxes represent active clauses. 
And-boxes require a more complex structure than or-boxes, as they store
information on which goals are active as well as on external and internal
variables associated with the clause. Figure \ref{fig:andbox} shows an and-box.

Access to the parent node is performed through the \texttt{parent}
pointer and through the \texttt{id\_alternative} field. The former
points back to the parent or-box. The later indicates to which
alternative the and-box belongs. Subgoal management requires knowing
the number of subgoals in the clause, \texttt{nr\_calls}.  Each
and-box maintains a \texttt{depth\_level} counter that is used to
classify variables.  The \texttt{locals} field maintain a list
of variables local to this and-box.  Variable control is discussed in
more detail in section~\ref{sec:localvar}.  A list of bindings to
external variables is accessed through the \texttt{externals}
field. The and-box may have suspended trying to bind some variables,
if so this is registered in the \texttt{suspended} field. If
predicates with side-effects are present in the goals of this and-box,
they are registered in the \texttt{side\_effects} field.  Last, each
subgoal or call requires separate information:
\begin{itemize}
\item a pointer to a corresponding or-box, \texttt{call}, initially
  empty; it is initialized when the call is open;
\item a pointer to the \texttt{locals} variables vector.  Each goal
  needs an entry to the {\tt locals} variables because the
  \emph{promotion} and compression rules may add other goals and other
  variables to the and-box.  Still, each goal needs to be able to
  identify its own local variables;

\item a pointer to the \texttt{code} for the subgoal;
\item \texttt{State} information says whether the goal is \texttt{ready}
  to enter execution, is \texttt{running}, or has entered the
  \texttt{success} or \texttt{fail} states. Goals may also be
  \texttt{suspended} or \texttt{waiting} on some variable, from which
  they will enter the \texttt{wake} state.
\end{itemize}

Initially each and-box has a fixed number of local variables. However,
the number of local variables in an and-box may increase since the
promotion rule allows one to promote local variables to a different
and-box. We discuss local variables next.

\subsection{Local Variables}
\label{sec:localvar}

\begin{figure}[htbp]
\centerline{
\includegraphics[width=12cm]{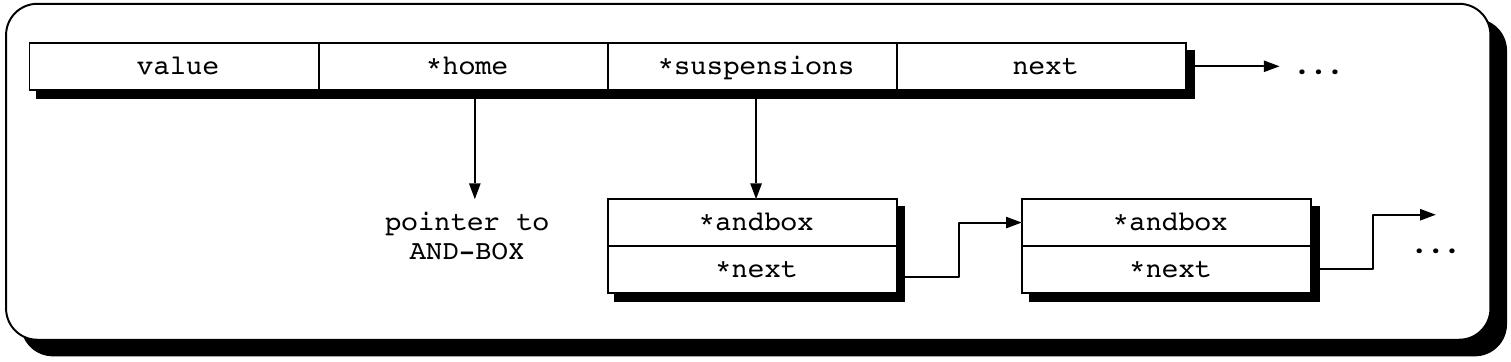}
}
\caption{Local Variables representation used in the BEAM: note that
  suspensions are explained in detail in figure~\ref{fig:suspensions}}
\label{fig:localvar}
\end{figure}

Every variable is a \emph{local variable} at some an-box; therefore it
is represented through the structure illustrated in
Figure~\ref{fig:localvar}. A local variable either belongs to a single
subgoal in an and-box, or it is shared among subgoals. The
\texttt{value} field stores the current working value for a variable.
Unbound variables are represented as self-referencing pointers.
Variables also maintain a list of and-boxes suspended on them, and
point back to their \texttt{home} and-box.

The \texttt{home} field of a variable structure points directly to its
original home and-box. However this field is not sufficient to completely
determine if a variable is local or not to an and-box.  


The BEAM detects
whether a variable is local to an and-box or not by having each
and-box associated with a
depth-counter that is then used to classify variables.  We can now recognize 
local variables as follows:

\begin{itemize} \item A variable occurring in an and-box 
  $\Delta$ is said to be local if the \texttt{depth} counter of the and-box $\Delta$
  equals the \texttt{depth} counter of the variable's \texttt{home}.
\item Otherwise the variable is said to be external to the and-box
  $\Delta$. \end{itemize}

\subsection{External Variables}

\begin{figure}[htbp]
\centerline{
\includegraphics{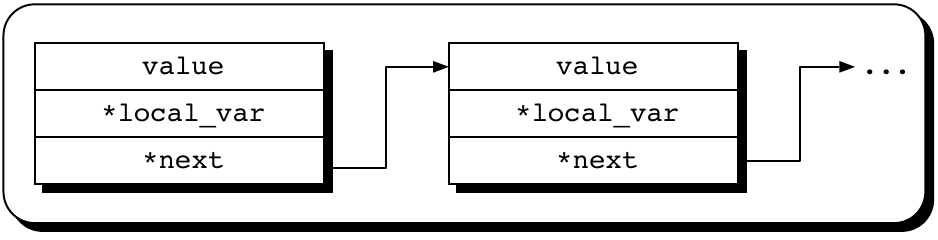}
}
\caption{External variables representation.}
\label{fig:externvar}
\end{figure}

Each and-box maintains a list of \emph{External Variables}, that is,
of bindings for variables older than the current and-box (see
Figure~\ref{fig:externvar}). Each such binding is represented as a
data structure that includes a pointer to the variable definition,
\texttt{local\_var}, and to its new value, \texttt{value}.  Whenever a
goal binds an external variable, the assignment is recorded both in
the current and-box as an external reference and at the local variable
itself. This way, whenever a descendant and-box wants to use the value
of the external reference, it can simply access the local variable. The
\texttt{external\_reference} data structure generalizes Prolog's
trail by allowing both the unwinding and the
rewinding of bindings performed in the current and-box. 
Our scheme for the external variables representation is 
very similar to the \emph{forward trail}~\cite{DSWarren84} used in the
SLG-WAM~\cite{Swift-PhD,Sagonas-98}.

\subsection{Suspension List}

\begin{figure}[htbp]
\centerline{
\includegraphics[width=12cm]{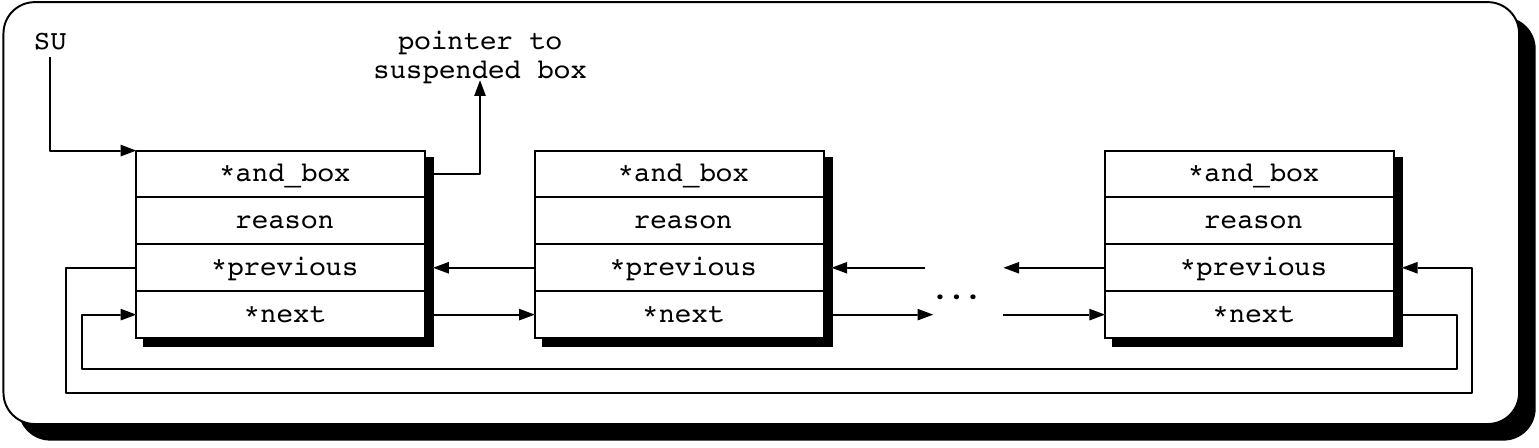}
}
\caption{Suspension list representation.}
\label{fig:suspensions}
\end{figure}

The suspension list is a doubly linked list that keeps information on
all suspended and-boxes (see figure~\ref{fig:suspensions}).  Each
entry in the list maintains a pointer to the suspended and-box, {\tt
and\_box}, and information on why the and-box suspended, {\tt reason}.
Goals may be suspended because they tried to bind external
variables. They can also be waiting for an event to occur. For
example, an I/O builtin may be waiting to be leftmost and an
arithmetic builtin may be waiting for a variable to become bound.

The AGENTS implementation uses one stack for suspended boxes and
another for woken boxes. In contrast, the BEAM uses the same list to
maintain information on suspended and woken and-boxes. The {\tt SU}
pointer marks the beginning of the suspension list (that can be
empty). Whenever an and-box suspends, an entry is added to the end of
the suspension list. If a suspended and-box receives a \texttt{wake}
signal, the and-box entry is moved to the beginning of the list. Thus,
if there are woken boxes, they are immediately accessed by the {\tt
  SU} pointer. Also note that we always want to work with woken boxes
before working with the suspended ones. By default, the BEAM chooses the leftmost
and-box in the And-Or Tree as the box to split first. The box is found
by depth-first search.

\section{Memory Management}
\label{sec:memory}


The EAM implements a flexible control strategy. Memory usage can
become a major concern in this case and the BEAM must carefully detect
the points at which to recover space.  As we show next, we have two
techniques to recover space: we can reuse space for pruned boxes and
we can garbage collect inaccessible data.

\subsection{Reusing Space in the And-Or Tree}
\label{sec:hybrid}

The \emph{Box Memory} must satisfy intensive requests for the creation
of and-boxes, or-boxes, local variables, external references, and
suspension lists. Objects are small and most, but not all, will have
short lifetimes.  Objects are created very frequently and
minimizing allocation and deallocation overheads is crucial. 

Unfortunately, the BEAM cannot recover space through
backtracking. Instead, it explicitly maintains liveness of data
structures, and relies on  a bucket allocation algorithm to
allocate space.

The BEAM is therefore able to recover all memory from boxes whenever
they fail or succeed.  Memory from failed boxes can obviously be
recovered since they do not add any knowledge to the computation.
Memory from successful boxes can also be recovered because the variable
unification rules guarantee that and-box variables do not reference
variables within the subtree rooted at this box, that is, younger box
variables can reference variables in upper boxes, but not the other
way around, as described further in section \ref{sec:unification}.

We have chosen this scheme because it has a low overhead and most
requests tend to vary among a relative small number of
sizes\cite{rslvsc_padl05}.

\subsection{Recovering Heap Space}

The algorithm used to reuse memory space in the \emph{Box Memory} will
not work for the Prolog terms stored in the \emph{Heap} because the BEAM releases memory eagerly,
and the terms in the \emph{Heap} tend to be very
small, causing fragmentation and leaving only small blocks available.
We could coalesce blocks to increase available block
size~\cite{detl94}, but the price would be an increase in
overheads. Instead, we have chosen to rely on a garbage collector to
compact the \emph{Heap Memory}.

We implemented a \emph{copying} garbage collector~\cite{gcbook,Bevemyr:1994:SEC}
for the BEAM: live data structures are copied to a new memory area and
the old memory area is released. The \emph{Heap} memory is divided
into two equal halves, growing in the same direction. The two halves
could not grow in the opposite direction because the BEAM uses YAP
builtins, and they expect the Heap to always grow upwards.  Therefore
we have a pre-defined limit-zone that, when reached, will activate the
garbage collection mechanism by setting the garbage collector flag.

The garbage collection flag is periodically checked by the And-Or Tree
manager to activate garbage collection.
Thus, the garbage collector starts by replicating the living data in
the root of the And-Or Tree and then follows a top-down-leftmost approach.


\subsection{Variable Allocation}

Variables are a major source of memory demand. In the initial
implementation of the BEAM, all variables were processed the same
way. Every and-box maintained a list of its local variables, and every
variable would be in some and-box. Let us refer to these variables as
\emph{permanent} variables. 
 
Processing all variables the same way has major drawbacks. Namely,
during the execution of a program there is a large portion of memory
that can be released only when the and-boxes fail or 
succeed. 

The complexity of this variable implementation can also harm system
performance.  Consider one of the main rules of the EAM, {\emph
Promotion}, used to promote the variables and constraints from
an and-box $\Delta$ to the nearest and-box $\Delta'$ above. $\Delta$
must be a single alternative to the parent or-box, as shown in Fig.
\ref{fig:beam-promotion}.

As in the original EAM promotion rule, promotion propagates results
from a local computation to the level above.  However, promotion in
the BEAM does not merge the two and-boxes because the structure of the
computation may be required to perform pruning as detailed in section 
\ref{sec:cut} ~\cite{rslvscfds_padl04}.

During the promotion of \emph{permanent} variables, the 
\emph{home} field of the variable structure needs to be updated so
that it points to the new and-box $\Delta'$. There is an overhead in
this operation since one must go through the list of all
\emph{permanent} variables of $\Delta$. Moreover if $\Delta'$ is
promoted later, the system will have to go through $\Delta'$ variables
including all that it has inherited during promotions. With
deterministic computations the list of \emph{permanent} variables can
grow very fast when promoting boxes, slowing down the BEAM.

\subsection{Classification of Variables at Compile and Run-Time}

Unfortunately, in general we do not know beforehand if we will need to
suspend on a variable. We propose a WAM-inspired scheme, the
{\bf BEAM-Lazy}. Following the WAM, variables that appear only in the body
of the clause or in queries are classified at compile time as
\emph{permanent} variables, meaning that all data-structures required
for suspension are created for them. Otherwise, variables are
classified at compile time as \emph{temporary}. 

As an example, consider the following clause of the nreverse procedure:
\begin{verbatim}
nreverse([X|L0],L) :- nreverse(L0,L1), concatenate(L1,[X],L).
\end{verbatim}

In this clause, \texttt{L1} is the only variable that is classified
as permanent at compilation time. The other variables are
classified as temporary. Thus, an
and-box for this clause will have one 
permanent variable and three temporary variables. 
Still, it may need to create two more
permanent variables, \texttt{X} and \texttt{L0}, when the
clause is called with the first argument as variable
(\texttt{unify\_var} when writing terms).

Temporary variables require less memory and improve performance
since we avoid managing the more complex structure of the
permanent variables. A second advantage from using
temporary variables is that they can be immediately released
after executing the clause body, unlike permanent variables
that can only be released when the and-box succeeds or fails. The BEAM
implements tail-recursion in the presence of deterministic
computation, so that temporary variables will be released before
calling the last subgoal.



\subsection{Variable Unification Rules}
\label{sec:unification}

The main consideration in implementing a unification algorithm that
supports both types of variables is that an and-box suspends only when
trying to bind \emph{permanent} variables external to the
and-box. 

There are three possible cases of variable-to-variable binding:
\begin{enumerate}
\item \emph{temporary variable} to \emph{permanent variable}: in this
  case the unification should make the \emph{temporary} variable refer to
  the \emph{permanent} variable. An immediate advantage is that the
  computation will not suspend. Unifying in the opposite
  direction would lead to an incorrect state.
  
\item \emph{temporary variable} to \emph{temporary variable}: the
  compiler ensures that a
  temporary variable is always bound to a permanent variable or a
  bound term, so this case will never occur.

\item \emph{permanent variable} to \emph{permanent variable}: the
  \emph{permanent} variable that has its home box at a lower level of the
  tree should always reference the \emph{permanent} variable that has its
  home box closer to the root of the tree.

\begin{figure}[htbp]
 \centerline{ \includegraphics{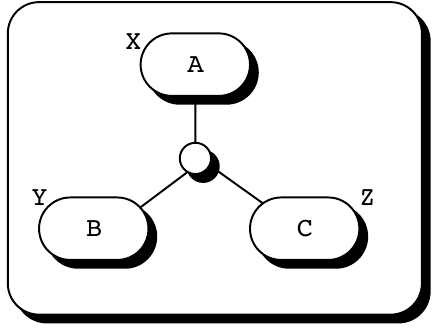} }
\caption{Binding two permanent variables.}
\label{fig:binding}
\end{figure}

Assume as an example the tree illustrated in
figure~\ref{fig:binding} with three and-boxes: {\tt A}, {\tt B}, and
{\tt C}. Each box contains a single permanent variable: {\tt X},{\tt
Y}, and {\tt Z}, respectively.  Assume that the computation is
processing the and-box {\tt B} and that it becomes necessary to unify the
variables {\tt X} and {\tt Y}.  If the variable {\tt Y} is made to
reference the variable {\tt X}, no suspension is necessary since the
variable {\tt Y} is local to the and-box {\tt B}.  Moreover, if the
and-box {\tt B} fails or if the computation continues to the and-box
{\tt C} no reset would be necessary in the  variable {\tt X}.  

\end{enumerate}

By following these unification rules one can often delay the suspension
of an and-box and thus delay application of the expensive splitting rule.


\section{The Emulator}
\label{sec:emulator}

The \emph{Emulator} is responsible for running WAM-like code in order
to perform unification and set up goals. The \emph{Emulator} executes
abstract machine instructions. The BEAM Emulator inherits most of
the WAM instructions and WAM registers. However, new instructions and
new registers are needed to cope with this rather different execution
model.

\subsection{Registers}
In a fashion similar to the WAM, the BEAM internal state is
saved in several registers:

\begin{itemize}
\item \texttt{PC}: Program Counter;
\item \texttt{H}: Top of Heap;
\item \texttt{S}: Structure pointer (points into the Heap);
\item \texttt{Mode}: controls whether unification is in read or write mode;
\item \texttt{X1,X2,...}: registers for temporary variables, also used as arguments registers;
\item \texttt{OBX}: pointer to the working or-box.
\item \texttt{ABX}: pointer to the working and-box.
\item \texttt{SU}: pointer to the list of suspended and-boxes.
\end{itemize}

Note that, except for the last three, the registers are inherited from
the WAM.  On the other hand, several of the WAM registers, such as {\tt
B}, {\tt ENV}, and {\tt HB},  are not needed in the BEAM
emulator as the BEAM does not implement backtracking. Instead
information is stored directly in the And-Or Tree.

\subsection{Abstract Machine Instructions}
\label{sec:absmachine}

Code for the BEAM abstract machine very closely follows the WAM.  The
BEAM abstract machine instructions include the WAM \texttt{get},
\texttt{put} and \texttt{unify} instructions, plus some novel control
instructions, that rely on the And-Or Tree Manager, described in
section~\ref{sec:tree_manager}. The main control instructions are:

\begin{description}
\item[\texttt{explore\_alternative i:}] explore the \texttt{i}th
  alternative for the current or-box. If there are more
  alternatives, create a new and-box. Otherwise, the parent and-box is reused for the alternative being
  executed (\emph{deterministic reduce and promote} optimization).  In
  both cases start executing
  the code for the unification of the arguments with the head of the
  goal.
  
\item[\texttt{prepare\_calls n:}] prepare the and-box to manage {\tt n}
  subgoals. Each subgoal record points to the start code for the
  call, and is initialized as \texttt{READY}, meaning that they are
  ready to be explored. If the and-box does not have external
  variables, execution is then passed to the \emph{And-Or Tree
  Manager} through \texttt{next\_call}. Otherwise, the
  and-box is marked as suspended, and execution enters the
  \emph{And-Or Tree Manager} through the \texttt{suspend} code.
  
\item[\texttt{call\_pred n:}] create one or-box with \texttt{n}
  branches, where \texttt{n} is the number of alternatives to
  \texttt{pred}. Each branch record points to the starting code of the
  corresponding alternative, and all branches are also initialized as
  {\tt READY}, meaning that they are ready to be explored. Execution
  is then passed to the \emph{And-Or Tree Manager}
  through the \texttt{next\_alternative} code.
  
\item[\texttt{proceed:}] return control from a clause
  to the \emph{Manager}. If the and-box does not have external
  variables, it has succeeded, and enters the \emph{And-Or Tree Manager}
  through the \texttt{success} module. Otherwise, the and-box is marked
  as suspended, and execution enters the \emph{And-Or Tree Manager}
  through the \texttt{suspend} module.
\end{description}

The major difference between the BEAM's \texttt{get}, \texttt{put},
and \texttt{unify} and the corresponding WAM instructions is that
whenever one of these instructions tries to bind an external variable,
an entry is added to the {\tt externals} field on the current and-box.
Note that in the WAM, during the unification of variables a check is also
done to determine if trailing is necessary. Thus, the BEAM {\tt
externals} field can be viewed as similar to the WAM trailing
mechanism.

\subsubsection{Compilation}

Compiling Prolog clauses to the BEAM abstract machine instructions is
very similar to WAM compilation. Figure~\ref{fig:beam_code}
illustrates an example of code generation.

\begin{figure}[htbp]
{\renewcommand{\baselinestretch}{1} 
\begin{verbatim}
         ancestor(X,Y):- parent(X,Y).
         ancestor(X,Z):- parent(X,Y), ancestor(Y,Z).

         parent(a,fa).                
         parent(a,ma).
-------------------------------------------------------------
                          ancestor/2

    explore_alternative  1       |    explore_alternative  2
    get_var          A1,Y1       |    get_var          A1,Y3
    get_var          A2,Y2       |    get_var          A2,Y1
    prepare_calls     1 L1       |    prepare_calls  2 L1 L2
 L1:                             | L1:
    put_val          A1,Y1       |    put_val          A1,Y3
    put_val          A2,Y2       |    put_val          A2,Y2
    call_pred     parent/2       |    call_pred     parent/2
                                 | L2:
                                 |    put_val          A1,Y2
                                 |    put_val          A2,Y1
                                 |    call_pred   ancestor/2
-------------------------------------------------------------
                           parent/2

    explore_alternative  1       |    explore_alternative  2
    get_atom         A1, a       |    get_atom         A1, a
    get_atom         A2,fa       |    get_atom         A2,ma
    procceed                     |    procceed 
                                 |
-------------------------------------------------------------
\end{verbatim} }
\caption{BEAM Abstract Machine Code for ancestor.}
\label{fig:beam_code}
\end{figure} 

Note that, unlike in the WAM, code for rules in the BEAM does not end
with an {\tt execute} instruction.  The BEAM abstract machine is goal
based. As such, the {\tt explore\_alternative i} instruction initializes the ith or-branch
by creating a new and-box in it. It is followed by a sequence of {\tt
  get} instructions that perform the head unification. Next, if the
clause is a fact, clause code terminates with the {\tt proceed}
instruction that decides whether the computation succeeds or whether
it should suspend (i.e., there are constraints on external variables),
entering the \emph{Manager} through the {\tt success} or the {\tt
  suspend} modules respectively.  If the clause is a rule, execution continues
with the {\tt
  prepare\_calls} instruction.  This instruction creates in
the and-box as many subgoals as calls.  Each subgoal is initialized to
point to the start code of each call (marked as {\tt L1}  and
\texttt{L2} in
figure~\ref{fig:beam_code}).  Then, the {\tt prepare\_calls} jumps to
the {\tt suspend} module if there are constraints on external
variables, or to the {\tt next\_call} port otherwise. Thus, it is up
to the \emph{Manager} to decide how and when to execute the calls.
The caller code is composed of a series of {\tt put} and possibly 
{\tt write} instructions followed by the {\tt call\_pred}
instruction. The {\tt call\_pred} instruction creates and initializes
an or-box with as many branches as the number of valid alternatives
(determined by the indexing on the first argument). Execution is then
passed to the Manager through the {\tt next\_alternative} port
that will decide which alternative to execute. By default, the
leftmost alternative is chosen.

\subsection{The And-Or Tree Manager}
\label{sec:tree_manager}

The \emph{And-Or Tree Manager} is the heart of our system. Its task is
to decide which rewrite rule should be applied to the current tree and
then execute it. The computational tree contains and-boxes and
or-boxes that can be in different states. The possible states for a
box are:
\begin{itemize}
\item \texttt{ready}: when a box is ready to start execution;
\item \texttt{running}: the box is already active and running;
\item \texttt{fail}: the box has failed;
\item \texttt{success}: the box has succeeded;
\item \texttt{suspended}: the box is suspended at some point;
\item \texttt{suspended-on-end}: the box is suspended and
  there are no more goals left to execute.  This is a special case
  of {\tt suspended}. The general case needs to continue the box
  execution. In this case we know that the execution is completed, so when
  the suspension is activated, the box can jump immediately to the {\tt
    success} state.
\item \texttt{awoken}: the box was suspended but has received a signal
  to be activated and can be restarted anytime.
\end{itemize}

The {\it And-Or Tree Manager} manages the states of and-boxes and
decides when to move boxes from one state to another.  

The {\it And-Or Tree Manager} is accessed through eight different
entry points:
\begin{description}

\item[\texttt{suspend:}] this routine adds the current and-box to the
  suspension list. Next, the routine clears all assignments saved in
  the list of the external variables.  Each external variable is also
  added to the suspension list included in the respective local
  variable.  After that, the \emph{And-Or Tree Manager} continues on
  \texttt{next\_alternative}.

\item[\texttt{success:}] this operation marks the current or-box as
  successful in its parent. The memory of the or-box is released.
  Next the parent and-box is checked. If all calls have 
  reached \texttt{success}, space for the and-box is reclaimed and
  the operation is reentered for the upper or-box (\emph{Success
  Propagation}).  Otherwise, execution enters the \texttt{next\_call}
  operation.
  
\item[\texttt{fail:}] this routine marks the current and-box as failed
  in its parent. All the assignments made by the and-box are removed,
  and space for the and-box is reclaimed. If all the alternatives for
  the parent or-box have failed, the operation is recursively called for
  the parent and-box (\emph{Failure Propagation}).  Otherwise, if
  there is exactly one more alternative, execution moves to   
  \texttt{unique\_alternative}. If there are several alternatives,
  execution continues at \texttt{next\_alternative}.
  
\item[\texttt{next\_call:}] this operation searches for the next
  non-suspended call in the current and-box.  If there is a ready call
  in the current and-box, \emph{Reduction} is applied by setting the
  \texttt{PC} to the start of the call's code, and then execution
  jumps to the \emph{Emulator}. Otherwise, if the and-box is not the
  root of the And-Or Tree, execution moves to
  \texttt{next\_alternative}.  If there is no ready call in the
  current and-box and if the current box is the root of the And-Or
  Tree, execution moves to the \texttt{select\_work} operation.
  
\item[\texttt{next\_alternative:}] this routine searches for the next
  non-suspended alternative in the current or-box to continue with the
  \emph{Reduction} of alternatives. If there is no such alternative,
  execution jumps to \texttt{next\_call}. Otherwise, if the
  alternative is in the \texttt{wake} state, execution moves to the
  {\tt wake} operation, else execution sets the \texttt{PC} to the
  code for the alternative code, and enters the \emph{Emulator}.
  
\item[\texttt{unique\_alternative:}] this operation applies a
  \emph{promotion} to the current and-box, since its parent or-box has a
  single alternative.  
  
  First, all external variables are checked, because after
  promotion some external variables may have become local. As a result,
  a wake signal is sent to all boxes suspended on this variable 
  (\emph{propagation}) that have their bindings promoted.  If during
  the promotion of external variables unification fails, execution
  moves to the \texttt{fail} operation. 

  Second, if external variables still
  exist after the promotion, the and-box remains suspended, and
  execution moves to \texttt{next\_call}. Otherwise, if the and-box is
  suspended and no more goals remain to execute, execution moves to
  \texttt{success}.

  Last, if goals are still left to execute, the \emph{And-Or Tree Manager}
  marks the and-box as {\it running} and continues its execution by
  entering the {\tt next\_call} operation.

\item[\texttt{wake:}] this operation chooses a suspended and-box that
  has received a wake up signal (\emph{propagation}). 

  First, all external
  variables are checked for changes: \emph{environment
    synchronization}. The \emph{environment synchronization} tests the
  compatibility of all the constraints imposed to
  external variables that are already bound. If unification fails,
  execution for the box jumps to \texttt{fail}. If unification
  succeeds for a variable, the and-box suspended on that variable can
  be deleted.  

  Last, If bindings to externals variables still exist,
  execution continues to \texttt{select\_work}.  If no more
  constraints on external variables are left, then the and-box enters
  \texttt{suspended\_on\_end}, and we can move immediately to {\tt
    success}. Otherwise, execution marks the and-box as {\it running}
  and continues its execution by entering the {\tt next\_call}
  operation.

\item[\texttt{select\_work:}] this operation looks for work in the 
  suspension list. If no extra work is available in the suspension
  list, execution will terminate. Otherwise, the \texttt{ABX} register
  is set to point to a box that is a candidate for \emph{splitting}. By default, the
  BEAM splits the leftmost suspended box.  After splitting, one of the
  resulting and-boxes is awakened, and its execution is restarted in
  \texttt{wake}.
\end{description}

\subsubsection{The Interaction Between the And-Or Tree Manager and the Emulator}

\begin{figure}[htbp]
\centerline{
\includegraphics{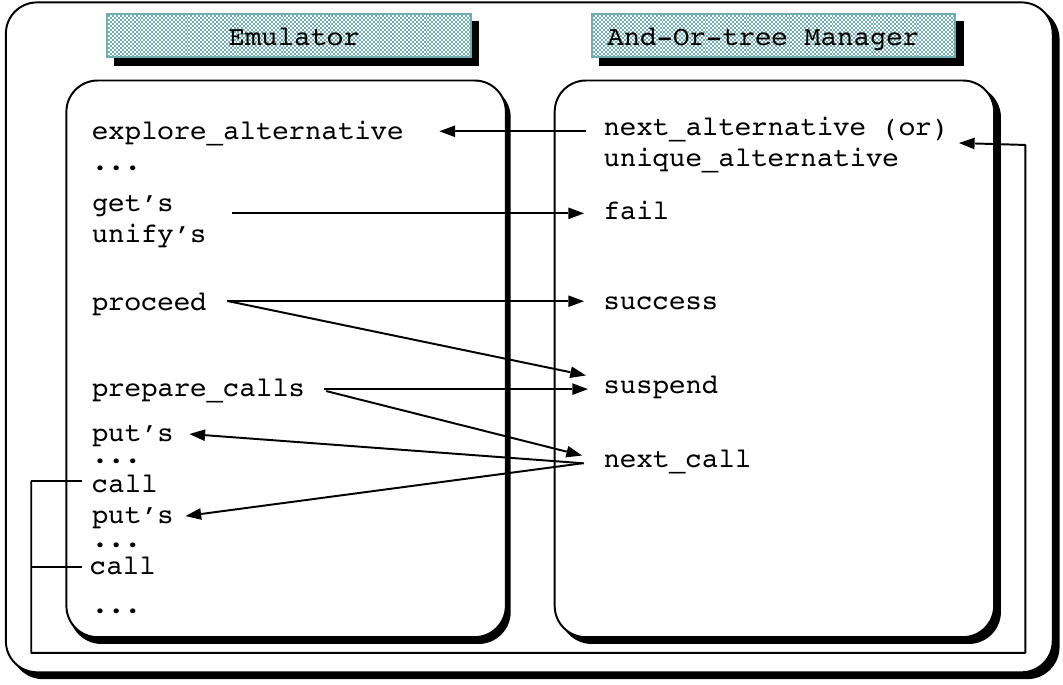}
}
\caption{Connecting the And-Or Tree Manager with the Emulator.}
\label{fig:conection}
\end{figure}

The \emph{And-Or Tree Manager} interacts with the \emph{Emulator} as
illustrated in figure \ref{fig:conection}.  In order to execute a
query, the \texttt{next\_alternative} first creates an or-box to store
all possible alternative clauses. An alternative is then chosen and
the execution passes to the \emph{Emulator}, through the
\texttt{explore\_alternative} instruction. Following this, execution
will run through a sequence of {\tt get} and possibly some {\tt unify}
instructions that implement the unification of the head arguments. If
this alternative is a fact then the emulator executes {\tt proceed}.
If there are assignments to external variables, this
instruction will move execution to the And-Or Tree Manager {\tt
suspend} operation. Otherwise execution moves to the {\tt success}
operation.

If the alternative is a rule, then instead of {\tt proceed} we have {\tt
prepare\_calls} followed by a sequence of {\tt put} and {\tt call\_pred}
instructions. The {\tt prepare\_calls} instruction creates an
and-box to store the calls and jumps to {\tt suspend} if 
there are assignments to external variables in the and-box, 
or to {\tt next\_call} otherwise.

The {\tt next\_call} operation chooses a call to execute, by default
the leftmost, and jumps to the \emph{Emulator} where it executes the
{\tt put} instructions followed by a {\tt call\_pred}. The {\tt
call\_pred} instruction will then jump to the {\tt next\_alternative}
operation in order to repeat the entire process.

We have so far considered the straightforward execution case.
Indeed, when running a normal program, the and-boxes will suspend when
constraining external variables. Thus, a computational state with all
and-boxes suspended is usual, and one must use the splitting rule
(\texttt{select\_work}) to create more deterministic work. The {\tt
select\_work} operation selects a candidate to split (by default the
leftmost suspended and-box). After splitting, the computation can
restart by waking one of the resulting and-boxes. The {\tt wake}
operation will then perform an environment check to determine if
the constraints being promoted are compatible with (possible)
constraints imposed by other and-boxes.


\section{Performance Analysis}
\label{sec:performance}

In this section we present the performance results of the prototype
BEAM system.
For the analysis of the BEAM performance we compare it 
with the following systems:
\begin{itemize}
\item {\bf SICStus Prolog 3.12.0}~\cite{sicstus}: is a
  state-of-the-art, ISO standard compliant, Prolog system developed at
  the SICS (the Swedish Institute of Computer Science). It is a commercial
  widely known system.  All benchmarks were executed using compiled
  emulated code.

\item {\bf YAP 5.0}~\cite{yap-optim}: is another state-of-the-art
  emulated Prolog system that was developed at University of Porto.
  This system is often regarded as the fastest Prolog system
  available for the PC Platform.
  
\item {\bf YAP 4.2}: is an older version of the YAP Prolog. 
BEAM was implemented on top of it.
  
\item {\bf Andorra-I v1.14}~\cite{VSCThesis}: is an implementation of
  the Basic Andorra Model that exploits or-parallelism and determinate
  dependent and-parallelism while fully supporting Prolog. We have
  used the sequential version for the comparison.  All benchmarks
  were pre-compiled by the Andorra-I preprocessor before execution.
  
\item {\bf AKL AGENTS v1.0}~\cite{JaMon92}: is a sequential Andorra
  Kernel Language implementation. The language was designed by Sverker
  Janson and Seif Haridi. AGENTS was developed by Johan Bevemyr and
  others, at SICS, Sweden.  This system follows an execution scheme
  that is similar to BEAM's but has the control intrinsic in the
  language. All benchmarks were rewritten to the AKL language before
  compiling and executing them on the Agents system.

\end{itemize}

We have used a representative group of well-known benchmarks. For each
benchmark we present the best execution time from a series of ten
runs. The runtime is presented for all systems in
milliseconds. Smaller benchmarks were run repeatedly. The
timings were measured running the benchmarks on an Intel Pentium
Mobile 1800Mhz (533Mhz FSB) with 2MB \emph{on chip} cache, equipped
with 1GB at 333Mhz DDR SDRAM and running Fedora Core 3. The BEAM was
configured with 64MB of \emph{Heap} plus 32MB of \emph{Box Memory}.
Benchmark code is available at
\url{http://www.dcc.fc.up.pt/~fds/rslopes}.


\subsection{The Benchmark Programs}

Table~\ref{tab:benchmarks} gives a small description of the benchmarks
used in this section. The selected group of benchmarks is composed by
well known test programs used within the Prolog community.

\begin{table}[htb]
\caption{The benchmarks.}
\centering
\begin{tabular}{ll} \hline
\multicolumn{2}{l}{\bf Deterministic:} \\ \hline
{\bf cal}      & last 10000 FoolsDays arithmetic benchmark.  \\
{\bf deriv}    & symbolically differentiates four functions of a single variable. \\
{\bf qsort}    & quick-sort of a 50-element list using difference lists. \\
{\bf serialise}& calculate serial numbers of a list. \\
{\bf reverse}  & smart reverse of a 1000-element list. \\
{\bf nreverse} & naive reverse of a 1000-element list. \\
{\bf kkqueens} & smart finder of the solutions for the n-queens problem. \\
{\bf tak}      & heavily recursive with lots of simple integer arithmetic. \\
\hline
\multicolumn{2}{l}{} \\
\hline
\multicolumn{2}{l}{\bf Non-Deterministic:} \\ \hline
{\bf ancestor} & query a static database. \\
{\bf houses}   & logical puzzle based on constraints. \\
{\bf query}    & finds countries with approximately equal population density.\\
{\bf zebra}    & logical puzzle based on constraints. \\
{\bf puzzle4x4}& finds a solution for a quadratic puzzle. \\ 
\hline
{\bf send}     & the SEND+MORE=MONEY puzzle. \\
{\bf scanner}  & a program to reveal the content of a box.\\
{\bf queens}   & finds safe placements of $n$-queens on $n*n$ chessboard. \\
{\bf check\_list}& list checker that verifies if duplicate elements exist. \\
{\bf ppuzzle}  & naive generation and test valid paths in a quadratic puzzle. \\
\hline
\end{tabular}
\label{tab:benchmarks}
\end{table}

The benchmarks are divided into two classes: deterministic and
non-deterministic. The non-deterministic benchmarks are further
subdivided into two groups: benchmarks that do not benefit from the
Andorra rule and benchmarks where the Andorra rule allows the search
space to be reduced.

\subsection{Performance on Deterministic Applications}

Table~\ref{tab:dtimes2} shows how the BEAM performs versus Andorra-I,
AGENTS, and the Prolog systems for deterministic applications. We use
SICStus Prolog as the reference system, so we give actual execution
times for SICStus and the relative time for the other systems. Neither
the BEAM nor AGENTS perform splitting, and Andorra-I always executes
deterministically. Prolog systems may create choice points.

\begin{table}[htbp] 
\caption{Deterministic benchmarks. SICStus Prolog is used as the
  reference system, with time given in milliseconds.}
\centering
\begin {tabular}{lcccccr} \hline 
  &  \bf SICStus & &  &  &\multicolumn{2}{c}{\bf YAP} \\
\raisebox{1.5ex}[0cm][0cm]{\bf Benchs.}  &
\multicolumn{1}{c}{\bf 3.12 } &
\raisebox{1.5ex}[0cm][0cm]{\bf BEAM}   &
\raisebox{1.5ex}[0cm][0cm]{\bf AGENTS}  &
\raisebox{1.5ex}[0cm][0cm]{\bf Andorra-I} &
 \multicolumn{1}{c}{\bf 4.2} &  \multicolumn{1}{c}{\bf 5.0} \\ \hline
\texttt{cal}          & 0.001   & 29\%    & 20\%   & 20\%      &  48\%  & 143\%  \\
\texttt{deriv}      & 0.010   & 31\%    &  8\%   & 33\%      &  72\%  & 128\% \\
\texttt{qsort}      & 0.045  & 23\%    & 14\%   & 24\%      &  88\%  & 102\%  \\
\texttt{serialise}  & 0.030  & 27\%    & 15\%   & 15\%      & 103\%  & 107\%   \\
\texttt{reverse\_1000}    & 0.050 & 27\%    & 12\%   & 15\%      &  42\%  & 116\% \\
\texttt{nreverse\_1000}  & 23   & 23\%    & 11\%   & 23\%      & 115\%  & 153\%   \\
\texttt{kkqueens}       & 30 & 27\%    & 20\%   & 20\%      &  58\%  & 158\%    \\
\texttt{tak}         & 16   & 31\%    & 33\%   & 23\%      & 160\%  & 133\%      \\
\hline\hline
\texttt{average}   &   & 27\%    & 17\%   & 22\%      &  86\%  & 130\%  \\
\hline
\end{tabular}
\label{tab:dtimes2}
\end{table}

The YAP and SICStus Prolog systems are recognized as some of the fastest
Prolog systems on the x86 architectures. The difference between Yap4.2
(on which the BEAM is based) and Yap5 shows that there is scope
for improvement even for Prolog systems. These improvements should
also benefit the BEAM.
Comparing with the BEAM, Yap5 is about 5 times faster than
the BEAM. SICStus Prolog and Yap4.2 are a bit less fast. This is quite
a good result for the BEAM, considering the extra complexity of the
Extended Andorra Model. 

The BEAM tends to perform better than the AGENTS especially on
tail-recursive computations. We believe this is because the BEAM has
special rules for performing tail-recursive computation that avoid
creating intermediate or-boxes and and-boxes. The results are
especially good for the BEAM considering that the BEAM does not need
any explicit control on these benchmarks. On the other hand, AGENTS
benefits from extra control to run the benchmarks deterministically.

Performance of the BEAM is very close to the performance of Andorra-I. 
Andorra-I beats BEAM on two benchmarks: \texttt{deriv} and
\texttt{qsort}.  This seems to depend on determinacy detection
performed by the Andorra-I preprocessor. Consider the following code
from the \texttt{qsort} benchmark:

\begin{verbatim}
   partition([X|L],Y,[X|L1],L2) :- X =< Y, partition(L,Y,L1,L2).
   partition([X|L],Y,L1,[X|L2]) :- X  > Y, partition(L,Y,L1,L2).
\end{verbatim}

Andorra-I classifies this code as deterministic, and never creates a
choice point. 
Unlike Andorra-I, BEAM does not have a pre-compilation with determinacy
analysis to classify this predicate as deterministic. Thus, the 
sophisticated determinacy code in Andorra-I can limit the overheads that
the BEAM has to go through by creating unnecessary and-boxes.
For better understanding these overheads that
BEAM suffers from, consider the two possible cases when running the {\tt
partition} predicate:
\begin{itemize}
\item {\tt X} $\leq$ {\tt Y}: the BEAM creates an or-box with two
  alternatives.  It performs the head unification for the first
  alternative and it succeeds executing the test comparing {\tt X} with
  {\tt Y}.  Execution then immediately suspends, because head
  unification generates bindings to variables external to the box.
  Execution then continues with the second alternative that will fail
  when comparing {\tt X} to {\tt Y}. This failure makes the first
  alternative the unique alternative in the or-box, and a promotion
  will occur allowing the suspended computation to resume.
  
\item {\tt X} $>$ {\tt Y}: the BEAM creates an or-box with two
  alternatives.  The first alternative will fail when comparing {\tt
    X} and {\tt Y}.  This failure makes the second alternative unique
  in its or-box. Promotion thus will occur, allowing the second
  alternative to run deterministically without suspending.
\end{itemize}

Concluding, the BEAM deterministic performance seems to be somewhat
better than AGENTS and equivalent to Andorra-I, although in some code
the BEAM still has greater overheads than Andorra-I.

\subsection{Performance on Non-Deterministic Applications}

Comparing different systems for non-deterministic benchmarks is
hard, since the search spaces may be quite different for Prolog, BEAM,
AGENTS, and Andorra-I. We will consider two classes of
non-deterministic applications. First, we consider applications where
the Andorra Model does not provide improve the search space. 
Note that in general one would not be particularly
interested in the BEAM for these applications: first, splitting is
very expensive and second, or-parallelism can also be quite
effectively exploited in Prolog. Next, we will consider examples where
the Andorra rule reduces, very significantly, the search space.

Table~\ref{tab:ndtimes2} shows a set of
five non-deterministic benchmarks. Again, we use SICStus Prolog as the
reference system. The number of splits for the BEAM
and AGENTS and the number of non-deterministic steps for Andorra-I 
for this set of benchmarks is presented in table~\ref{tab:ndsplits}.
We consider two versions of the BEAM. The default version delays
splitting until no other rules are applied. The \emph{ES} version uses
eager splitting. In this version splitting on producer goals is
performed immediately.  Producer goals are identified through the use
of an annotation inserted in the Prolog program.  Eager splitting makes
the BEAM computation rule closer to that of Prolog.  Note that to attain
good results with eager splitting, BEAM depends on the user to
identify the producer goals.  Ideally, we would prefer to use compile
time analysis instead.

\begin{table}[htbp] 
\caption{Non-deterministic benchmarks. SICStus Prolog is used as the
reference system with time given in milliseconds.}
\small
\centering
\begin {tabular}{lccccccr} \hline 
 & {\bf SICStus} & \multicolumn{2}{c}{\bf BEAM }  &  &  &\multicolumn{2}{c}{\bf YAP}  \\
\raisebox{1.5ex}[0cm][0cm]{\bf Benchs.} & \multicolumn{1}{c}{\bf 3.12}
& \multicolumn{1}{c} {\bf
Default} & \bf ~~ES~~ &
\raisebox{1.5ex}[0cm][0cm]{\bf AGENTS} & 
\raisebox{1.5ex}[0cm][0cm]{\bf AND.-I} &
\multicolumn{1}{c}{\bf 4.2} &  \multicolumn{1}{c}{\bf 5.0}  \\ \hline
\texttt{ancestor}  & 0.014 & N/A    & 38\%  &  6\%  & 16\%  & 139\% & 152\%\\ 
\texttt{houses}    & 0.37 & 79\%  & 82\%  & 86\%  & 46\%  & 132\% & 206\%  \\ 
\texttt{query}      & 0.30 &  3\%  & 12\%  &  2\%  & 29\%  &  85\% & 236\%  \\ 
\texttt{zebra}      & 7.50  & 33\%  & 79\%  & 40\%  & 32\%  &  97\% & 124\%  \\ 
\texttt{puzzle4x4} & 200 & 32\%  &  N/A    & 23\%  & 22\%  & 101\% & 121\%  \\ 
\hline\hline
\texttt{average}   & & 37\%  & 53\%  & 32\%  & 29\%  & 111\% & 168\%  \\ 
\hline
\end{tabular}
\label{tab:ndtimes2}
\end{table}

\begin{table}[htbp] 
\caption{Number of splits for BEAM/AGENTS and non-deterministic steps for Andorra-I.}
\centering
\begin {tabular}{lrrrr} \hline 
  & \multicolumn{2}{c}{\bf BEAM }  &  &  \\
\raisebox{1.5ex}[0cm][0cm]{\bf Benchs.} & \multicolumn{1}{c} {\bf Default}  &
\multicolumn{1}{c} {\bf ~~~~ES~~~~} &
\raisebox{1.5ex}[0cm][0cm]{\bf AGENTS} & 
\raisebox{1.5ex}[0cm][0cm]{\bf ANDORRA-I} \\ \hline
\texttt{ancestor}  & N/A    & 30  & 73     & 75  \\
\texttt{houses}    & 49     & 68  & 236    & 237    \\
\texttt{query}     & 624    & 624 & 624    & 626    \\
\texttt{zebra}     & 695    & 294 & 493    & 3,631  \\
\texttt{puzzle4x4} & 53,350 & N/A & 53,350 & 53,351 \\
\hline
\end{tabular}
\label{tab:ndsplits}
\end{table}



This set of benchmarks covers several major cases that can occur when
using eager splitting on BEAM.
\begin{itemize}
\item The \texttt{ancestor}
benchmark~\cite{gupta-warren91}, is one example where having producers
avoids
a situation where the EAM may loop.
\item The \texttt{houses} benchmark is an interesting case where, although
eager splitting increases the number of splits, performance still has
a slight improvement. This
example shows that splitting earlier with less data to copy may have
advantages in some cases.  Moreover, the BEAM has a lower number of
splits than AGENTS because the BEAM can delay splitting until \emph{success} or
\emph{failure propagation}, whereas AGENTS depends on guards.
\item Using eager splitting on the \texttt{query} benchmark does not change
the number of splits performed but has a huge improvement on system
performance.
\item The \texttt{zebra} benchmark is another demonstration of the impact of
eager-splitting in the EAM. In this example just defining the producer
dramatically cuts the search space and achieves much better
performance than AGENTS and Andorra-I. Moreover, the good execution
time when compared with Prolog indicates that the system actually
reduces the search space.
\item Finally, in the \texttt{puzzle} benchmark
the main goal suspends during the head unification and is forced to
perform splitting immediately. Thus, there is no early execution, and
no difference in using eager splitting.
\end{itemize}

\begin{figure}[htbp]
\centerline{
\includegraphics{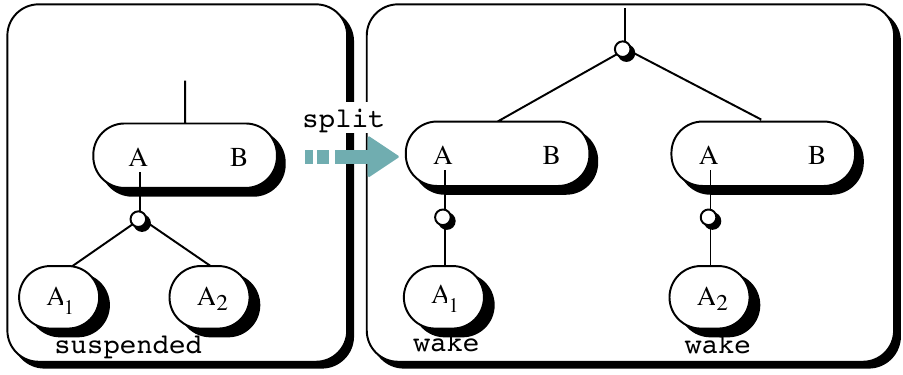}
}
\caption{Splitting effects.}
\label{fig:es_example}
\end{figure}

To better understand the effects of the splitting rule, and what
implications eager splitting can have on the computation tree, 
consider the example illustrated in figure~\ref{fig:es_example}. 

We assume two goals, the producer \emph{A} and the consumer \emph{B}.
The BEAM allows two methods to determine when to split on the goal
\emph{A}:
\begin{itemize}
\item {\bf default rule}: the split on \emph{A} will only be performed
  when all the computation on \emph{B} suspends. If \emph{A} is a
  producer, then there is a risk of having the EAM create
  speculative work on \emph{B} (in the worst case even leading to non-termination).
  Moreover, when splitting on \emph{A}, the entire And-Or Tree created on
  \emph{B} will be copied. This copying can be expensive and bring a large
  penalty to the execution time.

\item {\bf with eager splitting on A}: the split on \emph{A} will be
performed before starting execution of \emph{B}. The split
will be simpler since there will be no data associated with the
goal \emph{B} to replicate. The disadvantage is that this goal after
the splitting is totally unexplored in two and-boxes of the tree, and
thus there is duplication of work.
\end{itemize}
In general, eager splitting is appropriate when we expect that early
execution of the other goals will not constrain the producers. In
other words, early splitting should be pursued if we expect
\emph{splitting to be needed anyway}. In that case, early splitting
makes splitting much less expensive.

\subsubsection{Improving the Search Space}

The main benefit of the BEAM is in applications where we can
significantly improve the search space. Such applications may be pure
logic programs, or may be applications that take advantage of the
concurrency inherent to the Andorra Model. We consider five examples.
The \texttt{send\_more\_money} and the \texttt{scanner} benchmarks are  
well-known examples of declarative programs that perform badly in Prolog. A set of Prolog
benchmarks would not be complete without experimenting with a naive
solution to find the first solution for the \texttt{queens} problem.
And finally we consider two benchmarks that process lists, the
\texttt{check\_list} and the \texttt{ppuzzle}.  Each list element is a
pair with the form $p(X,Y)$ representing a position in an $n*n$
matrix. The {\tt check\_list} benchmark succeeds if an input list does
not hold duplicate elements. The \texttt{ppuzzle} generates all lists
with all possible combinations of the different elements in an $n*n$
matrix, and validates those that obey certain predefined
conditions.

Results are shown in table~\ref{tab:sstimes} and
table~\ref{tab:sssplits}.  The \texttt{send-more-money}, the
\texttt{scanner} and the \texttt{queens} benchmarks are quite
interesting because the BEAM without extra controls does not perform
more splits than AGENTS and it has slightly better performance.
Performance is three orders of magnitude faster than Prolog's.  These
benchmarks are also interesting in that they show a situation where
the more Prolog-like Andorra-I actually obtains the best results.
Although performing the same (or a few more) non-deterministic steps
as BEAM and AGENTS, Andorra-I is faster as choice-point manipulation
is more efficient than splitting.

\begin{table}[htbp]
\caption {Reduced search benchmarks (time in milliseconds).}
\centering
\begin {tabular}{lrrrr} \hline 
\bf Benchs.  & \bf BEAM  & \bf AGENTS & \bf ANDORRA-I & \bf YAP 5.0 \\
\hline
\texttt{send\_money}&  7 &  8  &  0.8 &   7,767   \\
\texttt{scanner    }& 20 & 39  &  3   & $>$12 hours \\ 
\hline
\texttt{queens-9}  &  2   &  8      & 0.9  &  16 \\
\texttt{queens-10} & 11   & 27      & 2.2  &  124 \\
\texttt{queens-11} &  7   & 19      & 1.3  &  893 \\ 
\texttt{queens-12} & 45   & 109     & 6.4  & 9,042 \\
\texttt{queens-13} & 23   & 58      & 3.4  & 93,343 \\
\texttt{queens-14} & 523  & 1,122   & 60   &1,175,535 \\
\texttt{queens-15} & 456  & 1,042   & 50   &15,287,308 \\  
\texttt{queens-16} &  3,958 & 8,363  & 396    & $>$12 hours \\ 
\texttt{queens-17} &  2,547 &  5,847 & 230    & $>$12 hours \\ 
\texttt{queens-18} & 21,891 &  47,113& 1,890  & $>$12 hours \\ 
\texttt{queens-19} &  1,572 &  3,796 &  120   & $>$12 hours \\ 
\texttt{queens-20} &138,799 & 302,048& 10,680 & $>$12 hours \\ 
\hline
\texttt{check\_list-8}  & 0.05 &  0.06 &      1,150  & 183    \\
\texttt{check\_list-9}  & 0.07 &  0.07 &      5,810  & 963    \\
\texttt{check\_list-10} & 0.09 &  0.09 &    209,140  & 34,910 \\
\texttt{check\_list-11} & 0.10 &  0.11 &  6,165,824  & 987,243\\
\texttt{check\_list-15} & 0.17 &  0.18 & $>$12 hours & $>$12 hours \\
\hline
\texttt{ppuzzle-A}      &  8   &   5   & 134,417     & $>$12 hours \\
\texttt{ppuzzle-B}      & 18   &  10   &$>$12 hours  & $>$12 hours \\
\texttt{ppuzzle-C} 1st  &  1.9 &  1.3  & 2,059,329   & $>$12 hours \\
\hline
\end{tabular}
\label{tab:sstimes}
\end{table}

\begin{table}[htbp]
\caption{Number of splits for BEAM/AGENTS and non-deterministic steps for Andorra-I.}
\centering
\begin {tabular}{lrrr} \hline
\bf Benchs.  & \bf BEAM  & \bf AGENTS & \bf ANDORRA-I \\
\hline \hline
\texttt{send\_money}& 277       & 277       &  325   \\
\texttt{scanner}    & 310       & 440       &  75    \\
\hline
\texttt{queens-9}   & 129       &  129      & 130   \\ 
\texttt{queens-10}  & 364       &  364      & 364   \\
\texttt{queens-11}  & 212       &  212      & 212   \\
\texttt{queens-12}  & 1,109     &  1,109    & 1,110 \\
\texttt{queens-13}  & 522       &  522      & 523   \\
\texttt{queens-14}  & 9,046     &  9,046    & 9,046 \\
\texttt{queens-15}  & 7,054     &  7,054    & 7,055 \\
\texttt{queens-16}  & 52,617    & 52,617    & 52,617    \\
\texttt{queens-17}  & 31,210    & 31,210    & 31,210    \\
\texttt{queens-18}  & 236,172   & 236,172   & 236,173   \\
\texttt{queens-19}  & 16,178    & 16,178    & 16,178    \\
\texttt{queens-20}  & 1,229,355 & 1,229,355 & 1,229,355 \\
\hline
\texttt{check\_list-8}  & 1  &  19 & 525,447    \\
\texttt{check\_list-9}  & 1  &  22 & 2,658,697  \\
\texttt{check\_list-10} & 1  &  26 & 95,365,524 \\
\texttt{check\_list-11} & 1  &  32 & - \\
\texttt{check\_list-15} & 1  &  64 & - \\
\hline
\texttt{ppuzzle-A}      & 86 &  86 & 64,994,853  \\
\texttt{ppuzzle-B}      &215 & 215 & - \\
\texttt{ppuzzle-C} 1st  & 29 &  29 & 751,567,244 \\
\hline
\end{tabular}
\label{tab:sssplits}
\end{table}

The \texttt{check\_list} and the \texttt{ppuzzle} benchmarks are 
examples where the EAM benefits from allowing non-deterministic goals
to execute as long as they do not bind external variables, as these
goals actually fail early. On the
other hand, Andorra-I is limited on this benchmark by the
non-determinism of the main predicates. On {\tt check\_list} Andorra-I
has a search space similar to Prolog's, while in the {\tt ppuzzle}
Andorra-I is better than Prolog, but still has a larger search
space than BEAM and AGENTS.  The difference seems to be that both the
BEAM and AGENTS benefit from early execution of the body of rules.

\section{Conclusions \& Related work}

We have presented the design and the implementation of the BEAM, a
system for the efficient execution of logic programs based on David H.\
D.\ Warren's work on the Extended Andorra Model with implicit control.
Our work was motivated by our interest in studying how the EAM with
implicit control can be effectively implemented and how it can perform
versus other execution strategies.  We believe the BEAM is a step towards
extending logic programming for applications where Prolog currently 
does not perform well. We believe that our results are quite
encouraging in this direction. 

Our approach contrasts with previous work in concurrent languages such
as AKL.  These are powerful concurrent languages that open up new
programming paradigms, but that also require users to invest in
sophisticated new programming frameworks.  In contrast, our first goal is 
to support a very flexible engine for the execution of logic
programs. The engine can then be controlled through several 
control primitives. 

The main contribution of this work is thus the design and
implementation of the BEAM. Further, our work in clarifying the EAM
and in designing the BEAM has shown a crisp separation between the
rewrite rules and control. We have tried to make this
separation clear in this presentation.  


In the future, we would like to explore different control
strategies over the basic rewrite-rules. Indeed control may be made
configurable, say, by using a specialized control language that can
generate specialized And-Or Tree managers. We believe that a major
contribution of the EAM is the exciting prospect of achieving
specialized control strategies for different types of logic programs.

The current BEAM prototype is available as part of the YAP Prolog
system distribution since release 5.1~\cite{DBLP:conf/iclp/Costa08}.
Although the BEAM is still a prototype, results are promising. The
BEAM appears as an alternative to run programs where standard Prolog
systems behave badly. Unlike AGENTS, the BEAM supports Prolog and
unlike Andorra-I it does not need pre-compilation analysis. Thus, the
BEAM is an excellent alternative for applications where
pre-compilation to Prolog may be expensive and difficult and where
queries with large search spaces are generated in rapid succession.

The BEAM prototype is currently being ported to the latest version of
YAP, with the new indexing algorithm~\cite{yap-jiti}, which should
further  improve BEAM performance. Currently, the BEAM only supports
Herbrand domain constraints. We plan to use YAP
attributed variable support to exloit non-Herbrand
constraints.  Ultimately, we aim at making the BEAM an extension of
Prolog systems that the user can  exploit towards maximum performance
in declarative applications.

We believe that the BEAM provides an excellent framework for novel
logic programming applications. We are particularly interested in
performance evaluation for automatically generated queries, say, the
ones that are found in Inductive Logic Programming~\cite{costa02query}.
In these applications, queries with large search spaces are generated
in rapid succession. Reducing the search space is fundamental, but
pre-compilation to Prolog may be expensive and difficult. We believe
that the advanced search features of the EAM can be most useful for
these applications.

\subsection*{Acknowledgments}
We would like to gratefully acknowledge the contributions we received
from Salvador Abreu, Gopal Gupta, Enrico Pontelli and Ricardo Rocha.
The work presented in this paper has been partially supported by
project HORUS (PTDC/EIA-EIA/100897/2008), LEAP
(PTDC/EIA-CCO/112158/2009), and funds granted to {\em
  LIACC} and \emph{CRACS \& INESC-Porto LA} through the {\em Programa de  Financiamento
  Plurianual, Funda\c{c}\~ao para a Ci\^encia e Tecnologia} and {\em
  Programa POSI}. Last, but not least, we would like to gratefully
acknowledge the anonymous referees for the major contributions that
they have given to this paper.

\bibliographystyle{acmtrans}
\bibliography{refs}

\begin{thebibliography}{}

\bibitem[\protect\citeauthoryear{Aggoun, Chan, Dufresne, Falvey, Grant, Herold,
  Macartney, Meier, Miller, Mudambi, Perez, van Rossum, Schimpf, Tsahageas, and
  de~Villeneuve}{Aggoun et~al\mbox{.}}{1995}]{Eclipse}
{\sc Aggoun, A.}, {\sc Chan, D.}, {\sc Dufresne, P.}, {\sc Falvey, E.}, {\sc
  Grant, H.}, {\sc Herold, A.}, {\sc Macartney, G.}, {\sc Meier, M.}, {\sc
  Miller, D.}, {\sc Mudambi, S.}, {\sc Perez, B.}, {\sc van Rossum, E.}, {\sc
  Schimpf, J.}, {\sc Tsahageas, P.~A.}, {\sc and} {\sc de~Villeneuve, D.~H.}
  1995.
\newblock {\em ECLiPSe 3.5 User Manual}.
\newblock ECRC.

\bibitem[\protect\citeauthoryear{Bevemyr and Lindgren}{Bevemyr and
  Lindgren}{1994}]{Bevemyr:1994:SEC}
{\sc Bevemyr, J.} {\sc and} {\sc Lindgren, T.} 1994.
\newblock A simple and efficient copying garbage collector for {Prolog}.
\newblock In {\em 6th International Symposium on Programming Language
  Implementation and Logic Programming, PLILP'94}. LNCS, vol. 844.
  Springer-Verlag, 88--101.

\bibitem[\protect\citeauthoryear{Bueno and Hermenegildo}{Bueno and
  Hermenegildo}{1992}]{bueher92}
{\sc Bueno, F.} {\sc and} {\sc Hermenegildo, M.~V.} 1992.
\newblock {An Automatic Translations Scheme from Prolog to the Andorra Kernel
  Language}.
\newblock In {\em {International Conference on Fifth Generation Computer
  Systems 1992}}. ICOT, Tokyo, Japan, 759--769.

\bibitem[\protect\citeauthoryear{Colmerauer}{Colmerauer}{1993}]{Col93}
{\sc Colmerauer, A.} 1993.
\newblock {The Birth of Prolog}.
\newblock In {\em The Second ACM-SIGPLAN History of Programming Languages
  Conference}. ACM, 37--52.

\bibitem[\protect\citeauthoryear{Detlefs, Dosser, and Zorn}{Detlefs
  et~al\mbox{.}}{1994}]{detl94}
{\sc Detlefs, D.}, {\sc Dosser, A.}, {\sc and} {\sc Zorn, B.} 1994.
\newblock Memory allocation costs in large {C} and {C++} programs.
\newblock {\em Softw. Pract. Exper.\/}~{\em 24,\/}~6, 527--542.

\bibitem[\protect\citeauthoryear{Gupta and Pontelli}{Gupta and
  Pontelli}{1997}]{Gupta:1997:EDD}
{\sc Gupta, G.} {\sc and} {\sc Pontelli, E.} 1997.
\newblock {Extended dynamic dependent And-parallelism in ACE}.
\newblock In {\em {PASCO} '97. Proceedings of the second international
  symposium on parallel symbolic computation, July 20--22, 1997, Maui, {HI}},
  {{ACM}}, Ed. ACM Press, New York, NY 10036, USA, 68--79.

\bibitem[\protect\citeauthoryear{Gupta, Pontelli, Ali, Carlsson, and
  Hermenegildo}{Gupta et~al\mbox{.}}{2001}]{par-survey}
{\sc Gupta, G.}, {\sc Pontelli, E.}, {\sc Ali, K.}, {\sc Carlsson, M.}, {\sc
  and} {\sc Hermenegildo, M.} 2001.
\newblock {Parallel Execution of Prolog Programs: A Survey}.
\newblock {\em ACM Transactions on Programming Languages and Systems\/}~{\em
  23,\/}~4, 1--131.

\bibitem[\protect\citeauthoryear{Gupta and Warren}{Gupta and
  Warren}{1991}]{gupta-warren91}
{\sc Gupta, G.} {\sc and} {\sc Warren, D. H.~D.} 1991.
\newblock {A}n {I}nterpreter for the {E}xtended {A}ndorra {M}odel.
\newblock Internal report, University of Bristol.

\bibitem[\protect\citeauthoryear{Hermenegildo and Greene}{Hermenegildo and
  Greene}{1991}]{andprolog}
{\sc Hermenegildo, M.~V.} {\sc and} {\sc Greene, K.} 1991.
\newblock {\&-P}rolog and its {P}erformance: {E}xploiting {I}ndependent
  {A}nd-{P}arallelism.
\newblock {\em New Generation Computing\/}~{\em 9,\/}~3,4, 233--257.

\bibitem[\protect\citeauthoryear{Hermenegildo and Nasr}{Hermenegildo and
  Nasr}{1986}]{HerNasr86}
{\sc Hermenegildo, M.~V.} {\sc and} {\sc Nasr, R.~I.} 1986.
\newblock {Efficient Management of Backtracking in AND-parallelism}.
\newblock In {\em Third International Conference on Logic Programming}. Number
  225 in Lecture Notes in Computer Science. Imperial College, Springer-Verlag,
  40--54.

\bibitem[\protect\citeauthoryear{Hill}{Hill}{1974}]{Hill74}
{\sc Hill, R.} 1974.
\newblock {LUSH-Resolution and its Completeness}.
\newblock Dcl memo 78, Department of Artificial Intelligence, University of
  Edinburgh.

\bibitem[\protect\citeauthoryear{I.S.Laboratory}{I.S.Laboratory}{2004}]{sicstus}
{\sc I.S.Laboratory}. 2004.
\newblock {\em SICStus Prolog user's manual, 3.12.0 Technical Report}.
\newblock Swedish Institute of Computer Science.

\bibitem[\protect\citeauthoryear{Janson and Haridi}{Janson and
  Haridi}{1991}]{JansonHSLP91}
{\sc Janson, S.} {\sc and} {\sc Haridi, S.} 1991.
\newblock {Programming Paradigms of the Andorra Kernel Language}.
\newblock In {\em {International Logic Programming Symposium, ILPS'91}}. MIT
  Press, 167--186.

\bibitem[\protect\citeauthoryear{Janson and Montelius}{Janson and
  Montelius}{1992}]{JaMon92}
{\sc Janson, S.} {\sc and} {\sc Montelius, J.} 1992.
\newblock {Design of a Sequential Prototype Implementation of the Andorra
  Kernel Language}.
\newblock {SICS Research Report}, {Swedish Institute of Computer Science}.

\bibitem[\protect\citeauthoryear{{Janson, Sverker}}{{Janson,
  Sverker}}{1994}]{SverkerThesis}
{\sc {Janson, Sverker}}. 1994.
\newblock {AKL -- A Multiparadigm Programming Language}.
\newblock Ph.D. thesis, {Uppsala University}.

\bibitem[\protect\citeauthoryear{Jones and Lins}{Jones and Lins}{1996}]{gcbook}
{\sc Jones, R.} {\sc and} {\sc Lins, R.} 1996.
\newblock {\em Garbage Collection: Algorithms for Automatic Dynamic Memory
  Management}.
\newblock John Wiley and Sons.
\newblock Reprinted February 1997.

\bibitem[\protect\citeauthoryear{Lloyd}{Lloyd}{1987}]{Lloyd87}
{\sc Lloyd, J.~W.} 1987.
\newblock {\em Foundations of Logic Programming\/}, second ed.
\newblock Springer-Verlag.

\bibitem[\protect\citeauthoryear{Lopes and {Santos Costa}}{Lopes and {Santos
  Costa}}{2005}]{rslvsc_padl05}
{\sc Lopes, R.} {\sc and} {\sc {Santos Costa}, V.} 2005.
\newblock Improving memory usage in the beam.
\newblock In {\em 7th International Symposium on Pratical Aspects of
  Declarative Languages, PADL'05}, {M.~Hermenegildo} {and} {D.~Cabeza}, Eds.
  LNCS, vol. 3350. Springer-Verlag, 143--157.

\bibitem[\protect\citeauthoryear{Lopes, {Santos Costa}, and Silva}{Lopes
  et~al\mbox{.}}{2001}]{rslvscfds_padl01}
{\sc Lopes, R.}, {\sc {Santos Costa}, V.}, {\sc and} {\sc Silva, F.} 2001.
\newblock A novel implementation of the extended andorra model.
\newblock In {\em 3rd International Symposium on Pratical Aspects of
  Declarative Languages, PADL'01}, {I.~V. Ramakrishnan}, Ed. LNCS, vol. 1990.
  Springer-Verlag, 199--213.

\bibitem[\protect\citeauthoryear{Lopes, {Santos Costa}, and Silva}{Lopes
  et~al\mbox{.}}{2003a}]{rslvscfds_iclp03}
{\sc Lopes, R.}, {\sc {Santos Costa}, V.}, {\sc and} {\sc Silva, F.} 2003a.
\newblock On deterministic computations in the extended andorra model.
\newblock In {\em 19th International Conference on Logic Programming, ICLP03},
  {C.~Palamidessi}, Ed. LNCS, vol. 2916. Springer-Verlag, 407--421.

\bibitem[\protect\citeauthoryear{Lopes, {Santos Costa}, and Silva}{Lopes
  et~al\mbox{.}}{2003b}]{rslvscfds_clps03}
{\sc Lopes, R.}, {\sc {Santos Costa}, V.}, {\sc and} {\sc Silva, F.} 2003b.
\newblock On the beam implementation.
\newblock In {\em 11th Portuguese Conference on Artificial Intelligence, EPIA
  2003}, {F.~M. Pires} {and} {S.~Abreu}, Eds. LNCS, vol. 2902. Springer-Verlag,
  131--135.

\bibitem[\protect\citeauthoryear{Lopes, {Santos Costa}, and Silva}{Lopes
  et~al\mbox{.}}{2004}]{rslvscfds_padl04}
{\sc Lopes, R.}, {\sc {Santos Costa}, V.}, {\sc and} {\sc Silva, F.} 2004.
\newblock Prunning in the extended andorra model.
\newblock In {\em 6th International Symposium on Pratical Aspects of
  Declarative Languages, PADL'04}, {B.~Jayaraman}, Ed. LNCS, vol. 3057.
  Springer-Verlag, 120--134.

\bibitem[\protect\citeauthoryear{Montelius and Ali}{Montelius and
  Ali}{1995}]{Penny}
{\sc Montelius, J.} {\sc and} {\sc Ali, K. A.~M.} 1995.
\newblock {An And/Or-Parallel Implementation of AKL}.
\newblock {\em New Generation Computing\/}~{\em 13,\/}~4, 31--52.

\bibitem[\protect\citeauthoryear{Montelius and Magnusson}{Montelius and
  Magnusson}{1997}]{PennySimICS}
{\sc Montelius, J.} {\sc and} {\sc Magnusson, P.} 1997.
\newblock Using {SIMICS} to evaluate the {Penny} system.
\newblock In {\em International Logic Programming Symposium, {ILPS}'97},
  {J.~Ma{\l}uszy{\'n}ski}, Ed. MIT Press, Cambridge, 133--148.

\bibitem[\protect\citeauthoryear{Palmer and Naish}{Palmer and
  Naish}{1991}]{PalNai91}
{\sc Palmer, D.} {\sc and} {\sc Naish, L.} 1991.
\newblock {NUA-Prolog: an Extension to the WAM for Parallel Andorra}.
\newblock In {\em 8th International Conference on Logic Programming, ICLP'91},
  {K.~Furukawa}, Ed. MIT Press.

\bibitem[\protect\citeauthoryear{Sagonas}{Sagonas}{1996}]{Sagonas-PhD}
{\sc Sagonas, K.} 1996.
\newblock {The SLG-WAM: A Search-Efficient Engine for Well-Founded Evaluation
  of Normal Logic Programs}.
\newblock Ph.D. thesis, Department of Computer Science, State University of New
  York, Stony Brook, USA.

\bibitem[\protect\citeauthoryear{Sagonas and Swift}{Sagonas and
  Swift}{1998}]{Sagonas-98}
{\sc Sagonas, K.} {\sc and} {\sc Swift, T.} 1998.
\newblock {An Abstract Machine for Tabled Execution of Fixed-Order Stratified
  Logic Programs}.
\newblock {\em ACM Transactions on Programming Languages and Systems\/}~{\em
  20,\/}~3 (May), 586--634.

\bibitem[\protect\citeauthoryear{Sagonas, Swift, Warren, Freire, and
  Rao}{Sagonas et~al\mbox{.}}{1997}]{xsb-manual}
{\sc Sagonas, K.}, {\sc Swift, T.}, {\sc Warren, D.~S.}, {\sc Freire, J.}, {\sc
  and} {\sc Rao, P.} 1997.
\newblock The {XSB} pro\-grammer's manual.
\newblock Tech. rep., State University of New York at Stony Brook.
\newblock Available at \url{http://www.cs.sunysb.edu/~sbprolog}.

\bibitem[\protect\citeauthoryear{{Santos Costa}}{{Santos
  Costa}}{1993}]{VSCThesis}
{\sc {Santos Costa}, V.} 1993.
\newblock {Compile-Time Analysis for the Parallel Execution of Logic Programs
  in Andorra-I}.
\newblock Ph.D. thesis, University of Bristol.

\bibitem[\protect\citeauthoryear{Santos~Costa}{Santos~Costa}{1999}]{yap-optim}
{\sc Santos~Costa, V.} 1999.
\newblock {Optimising Bytecode Emulation for Prolog}.
\newblock In {\em {International Conference Principles and Practice of
  Declarative Programming, PPDP'99}}. Springer-Verlag, LNCS 1702, 261--267.

\bibitem[\protect\citeauthoryear{Santos~Costa}{Santos~Costa}{2008}]{DBLP:conf/iclp/Costa08}
{\sc Santos~Costa, V.} 2008.
\newblock The life of a logic programming system.
\newblock In {\em 24th International Conference on Logic Programming, ICLP
  2008}, {M.~G. de~la Banda} {and} {E.~Pontelli}, Eds. LNCS, vol. 5366.
  Springer-Verlag, 1--6.

\bibitem[\protect\citeauthoryear{{Santos Costa}, Damas, Reis, and
  Azevedo}{{Santos Costa} et~al\mbox{.}}{2000}]{yap}
{\sc {Santos Costa}, V.}, {\sc Damas, L.}, {\sc Reis, R.}, {\sc and} {\sc
  Azevedo, R.} 2000.
\newblock {\em {YAP User's Manual}}.
\newblock Universidade do Porto.
\newblock available at \url{http://www.dcc.fc.up.pt/~vsc/Yap}.

\bibitem[\protect\citeauthoryear{Santos~Costa, Sagonas, and Lopes}{Santos~Costa
  et~al\mbox{.}}{2007}]{yap-jiti}
{\sc Santos~Costa, V.}, {\sc Sagonas, K.}, {\sc and} {\sc Lopes, R.} 2007.
\newblock Demand-driven indexing of prolog clauses.
\newblock In {\em 23rd International Conference on Logic Programming, ICLP'07},
  {V.~Dahl} {and} {I.~Niemel\"a}, Eds. LNCS, vol. 4670. Springer, 305--409.

\bibitem[\protect\citeauthoryear{Santos~Costa, Srinivasan, Camacho, Blockeel,
  Demoen, Janssens, Struyf, Vandecasteele, and Van~Laer}{Santos~Costa
  et~al\mbox{.}}{2003}]{costa02query}
{\sc Santos~Costa, V.}, {\sc Srinivasan, A.}, {\sc Camacho, R.}, {\sc Blockeel,
  H.}, {\sc Demoen, B.}, {\sc Janssens, G.}, {\sc Struyf, J.}, {\sc
  Vandecasteele, H.}, {\sc and} {\sc Van~Laer, W.} 2003.
\newblock {Query Transformations for Improving the Efficiency of ILP Systems}.
\newblock {\em Journal of Machine Learning Research\/}~{\em 4}, 465--491.

\bibitem[\protect\citeauthoryear{{Santos Costa}, Warren, and Yang}{{Santos
  Costa} et~al\mbox{.}}{1991a}]{ppopp}
{\sc {Santos Costa}, V.}, {\sc Warren, D. H.~D.}, {\sc and} {\sc Yang, R.}
  1991a.
\newblock {Andorra-I: A Parallel Prolog System that Transparently Exploits both
  And- and Or-Parallelism}.
\newblock In {\em {3rd ACM SIGPLAN Symposium on Principles \& Practice of
  Parallel Programming, PPOPP'91}}. ACM press, 83--93.
\newblock SIGPLAN Notices vol 26(7), July 1991.

\bibitem[\protect\citeauthoryear{{Santos Costa}, Warren, and Yang}{{Santos
  Costa} et~al\mbox{.}}{1991b}]{iclp91p}
{\sc {Santos Costa}, V.}, {\sc Warren, D. H.~D.}, {\sc and} {\sc Yang, R.}
  1991b.
\newblock {The Andorra-I Preprocessor: Supporting full Prolog on the Basic
  Andorra model}.
\newblock In {\em 8th International Conference on Logic Programming, ICLP'91}.
  MIT Press, 443--456.

\bibitem[\protect\citeauthoryear{Smolka}{Smolka}{1995}]{Oz}
{\sc Smolka, G.} 1995.
\newblock The {Oz} programming model.
\newblock In {\em Computer Science Today}, {J.~van Leeuwen}, Ed. LNCS, vol.
  1000. Springer-Verlag, Berlin, 324--343.

\bibitem[\protect\citeauthoryear{Swift}{Swift}{1994}]{Swift-PhD}
{\sc Swift, T.} 1994.
\newblock {Efficient Evaluation of Normal Logic Programs}.
\newblock Ph.D. thesis, Department of Computer Science, State University of New
  York, Stony Brook, USA.

\bibitem[\protect\citeauthoryear{Ueda}{Ueda}{2002}]{ghc02}
{\sc Ueda, K.} 2002.
\newblock {A Pure Meta-interpreter for Flat GHC, a Concurrent Constraint
  Language}.
\newblock In {\em Computational Logic: Logic Programming and Beyond}. LNCS,
  vol. 2407. Springer, 138--161.

\bibitem[\protect\citeauthoryear{Ueda and Morita}{Ueda and
  Morita}{1990}]{UedaMor}
{\sc Ueda, K.} {\sc and} {\sc Morita, M.} 1990.
\newblock {A New Implementation Technique for Flat GHC}.
\newblock In {\em 7th International Conference on Logic Programming, ICLP'89}.
  {MIT} Press, 3--17.

\bibitem[\protect\citeauthoryear{Warren}{Warren}{1983}]{Warren83}
{\sc Warren, D. H.~D.} 1983.
\newblock {An Abstract Prolog Instruction Set}.
\newblock Technical Note 309, SRI International.

\bibitem[\protect\citeauthoryear{Warren}{Warren}{1988}]{War88}
{\sc Warren, D. H.~D.} 1988.
\newblock The {Andorra} model.
\newblock Presented at Gigalips Project workshop, University of Manchester.

\bibitem[\protect\citeauthoryear{Warren}{Warren}{1989}]{War89}
{\sc Warren, D. H.~D.} 1989.
\newblock {Extended Andorra model}.
\newblock PEPMA Project workshop, University of Bristol.

\bibitem[\protect\citeauthoryear{Warren}{Warren}{1990}]{War90}
{\sc Warren, D. H.~D.} 1990.
\newblock {The Extended Andorra Model with Implicit Control}.
\newblock Presented at ICLP'90 Workshop on Parallel Logic Programming, Eilat,
  Israel.

\bibitem[\protect\citeauthoryear{Warren}{Warren}{1984}]{DSWarren84}
{\sc Warren, D.~S.} 1984.
\newblock {Efficient Prolog Memory Management for Flexible Control Strategies}.
\newblock In {\em International Logic Programming Symposium, ILPS'84}. Atlantic
  City, IEEE Computer Society, 198--203.

\end{thebibliography}

\label{lastpage}
\end{document}